\documentclass[lineno]{jfm}

\usepackage{graphicx}
\usepackage{epstopdf,epsfig}
\usepackage{newtxtext}
\usepackage{newtxmath}
\usepackage{natbib}
\usepackage{hyperref}
\hypersetup{
    colorlinks = true,
    urlcolor   = blue,
    citecolor  = black,
}

\newcommand{\RomanNumeralCaps}[1]
\linenumbers
\newcommand{\del}{\partial}

\newcommand{\doublecontract}{\mathrel{\substack{\bullet \\ \cdot}}}
\usepackage{comment}
\DeclareMathOperator\arctanh{arctanh}
\DeclareMathAlphabet{\mathsfbi}{OT1}{\sfdefault}{bx}{sl}
\DeclareMathVersion{sfletters}
\SetSymbolFont{letters}{sfletters}{OML}{ntxsfmi}{b}{it}
\makeatletter
\newcommand{\mathsfbilow}[1]{%
  \text{\mathversion{sfletters}$\m@th#1$}%
}
\makeatother

\title{Hydrodynamic particle interactions in linear and radial viscosity gradients}

\author{Sebastian Ziegler\aff{1},
 \and Ana-Sun\v{c}ana Smith\aff{1,2}\corresp{\email{smith@physik.fau.de}}}

\affiliation{\aff{1}Friedrich-Alexander-University Erlangen-Nürnberg (FAU), Faculty of Science, Department of Physics, PULS Group, Interdisciplinary Center for Nanostructured Films (IZNF), Cauerstrasse 3, 91058 Erlangen, Germany 
\aff{2}Group for Computational Life Sciences, Division of Physical Chemistry, Ruđer Bošković Institute, Bijenička cesta 54, 10000 Zagreb, Croatia}

\begin{document}
\maketitle

\begin{abstract} 
We present a versatile perturbative calculation scheme to determine the mobility matrix for two and more particles in a low Reynolds number fluid with spatially variant viscosity. 
Assuming an asymptotic non-constant viscosity perturbation superimposed on a constant viscosity background, we exploit the Lorentz reciprocal theorem and a reflection method to obtain the leading order correction to the mobility matrix. 
We apply our approach firstly to an interface-like linear viscosity gradient with an extension much larger than the length scale of the particle separation. We find that the viscosity gradient gives rise to additional long-range flow fields, for both particle translation and rotation, which decay by one order slower than their constant-viscosity counterparts. Also, we reveal that the relative positioning of two interacting particles in finite-size gradients, a natural precondition to avoid pathological negative viscosity, affects the mobilities of and interaction between both particles. 
Secondly, we apply our result to two interacting particles with temperatures different from that of the surrounding fluid, which alters both the fluid temperature and viscosity field. We find that both, the self-mobilities of the particles as well as their interactions, increase for hot particles and decrease for cold particles.
\end{abstract}

\section{Introduction}
Hydrodynamic interactions between particles of approximately spherical shape are of great importance for many areas in research and industry, including colloid dynamics \citep{Rex2008}, polymers \citep{Kirkwood1948}, deposition processes \citep{Dabros1989} and active systems such as microswimmers \citep{GolestanianAjdari2008} and Janus particles \citep{Oyama2016}. 
Due to the small length scales involved, the corresponding hydrodynamic flows are dominated by viscous friction rather than by fluid inertia. In this regime, the Navier-Stokes equations reduce to the Stokes equations which can be solved analytically at spatially constant viscosity for many geometries. 
Since viscosity gradients are abundant in both living and non-living environments, such as bacterial biofilms \citep{Wilking2011}, mucus layers \citep{Swidsinski2007} or interfaces between fluids of different viscosities \citep{Qiu2011}, it is natural to ask how the hydrodynamic interactions change in variant viscosities. The simplest example of such an environment is a linear viscosity gradient, i.e. a region in the fluid where the viscosity changes linearly with the position in space. Another experimentally relevant example is the viscosity perturbation induced by temperature contrasts between the interacting particles and the surrounding fluid \citep{Oppenheimer2016, Kroy2016}.
In both cases, results have already been obtained for the mobilities of single particles \citep{Datt2019a, Oppenheimer2016}, while only little is known for the interaction between several particles. 

In their seminal paper on microswimmer behaviour in a viscosity gradient \citep{Liebchen2018}, Liebchen et al estimated an upper bound to the amplitude of the leading order correction for the Oseen tensor within an infinite linear viscosity gradient but neglected those terms in their subsequent analysis of swimmer behaviour in a viscosity gradient. Laumann \& Zimmermann subsequently calculated the first order correction to the Oseen tensor in an infinitely large linear viscosity gradient \citep{Laumann2019}. However, to the best of our knowledge, no explicit results are yet available in the literature for the corrections to the full mobility matrix of two particles in non-homogeneous environments. Also, the studies mentioned did assume viscosity gradients which extend infinitely, although this inevitably introduces negative viscosity in some parts of the fluid. To avoid this, we explicitly consider viscosity gradients that, while being larger than the separation between both particles by orders of magnitude, are yet of finite size, and thus avoid negative fluid viscosity. 
For the second application of particles with a temperature difference with respect to the surrounding fluid, no results are yet available for the interactions between particles. 

We fill this gap, providing the leading order correction to the interaction of particles at large separations for both a large viscosity gradient as well as for viscosity perturbations due to the particle temperature. In the former case, we surprisingly find that the correction terms, which are to leading order linear in the viscosity gradient, scale constant in the distance between both particles, contrasting with the Oseen tensor which decays with the first power of the inverse distance. The correction to the flow field induced by a rotating particle is found to scale with the first power of the inverse distance to the particle, while the constant viscosity flow scales with the second power.
Also, we discover that an asymmetry in the domain of the viscosity gradient introduces an additional term which has a non-negligible impact on both the self-mobility as well as the interaction between the two particles. 
For particles with a temperature difference to the surrounding fluid, the interaction is corrected by a term proportional to the second inverse power of the particle separation. Even more prominently, also the self-mobilities of both particles are altered due to the presence of another particle.

The following work is structured as follows: We firstly introduce the problem of calculating the mobility matrix in non-constant viscosity in Section \ref{ch:problemFormulation}, before we give a short derivation of the perturbative first order correction to the mobility matrix following \citet{Oppenheimer2016} (Section \ref{ch:perturbationAnalysis}). 
We then employ a reflection method in order to expand the correction in the inverse separation of both particles, deferring the explicit calculation of the first order terms to the Appendix.
In Section \ref{ch:linearGradient}, we present results for two particles in a linear viscosity gradient of finite size, where we focus firstly on symmetric placement of particles in the gradient (Section \ref{ch:symmetricLinearGradient}) and second describe the additional terms arising when the particles are placed asymmetrically in the gradient (Section \ref{ch:asymmetricLinearGradient}).
As a second application, we focus on the mobility of a two particle system with a temperature contrast between the particles and the fluid at infinity (Section \ref{ch:hotParticles}). Section \ref{ch:conclusion} concludes the paper.

\section{Formulation of the problem}
\label{ch:problemFormulation}
We aim to calculate the mobility matrix for two spherical particles of equal radius $a$ with relative distance $s$ between the centres of both particles, which are suspended in a fluid with position-dependent viscosity $\eta(\boldsymbol{r})$. We assume that the fluid viscosity is given by
\begin{equation}
\eta(\boldsymbol{r}) = \eta^{(0)} + \epsilon \eta^{(1)}(\boldsymbol{r}), 
\end{equation}
with $\eta^{(0)}$ a constant reference viscosity, $\epsilon < 1$ the perturbation parameter and $\boldsymbol{r}$ the position in the fluid. Upper indices in brackets correspond to orders in $\epsilon$. We assume that the viscosity perturbation is at most of the order of $\eta^{(0)}$, such that $\epsilon$ corresponds to the ratio of the magnitudes of viscosity perturbation and the constant background viscosity. The Reynolds number \citep{Purcell1977} associated with the particles is assumed to be zero, such that the fluid is described by the incompressible Stokes equations \citep[p. 9]{KimKarilla1991}
\begin{equation}
\bnabla \boldsymbol{\cdot} \mathsfbilow{\sigma} (\boldsymbol{r}) = 0, \ \ \bnabla \boldsymbol{\cdot} \boldsymbol{u} (\boldsymbol{r}) = 0, 
\label{eq:StokesEquation}
\end{equation}
with $\boldsymbol{u} (\boldsymbol{r})$ the fluid velocity and $\mathsfbilow{\sigma} (\boldsymbol{r})$ the stress tensor defined as
\begin{equation}
\mathsfbilow{\sigma} (\boldsymbol{r}) = - p(\boldsymbol{r}) \mathsfbi{I} + 2 \eta(\boldsymbol{r}) \mathsfbi{E} (\boldsymbol{r}).
\end{equation}
Here, $p (\boldsymbol{r})$ denotes the pressure in the fluid, $\mathsfbi{E}(\boldsymbol{r}) = 1/2 \left(\bnabla \boldsymbol{u} (\boldsymbol{r}) + (\bnabla \boldsymbol{u}(\boldsymbol{r}))^T \right)$ the strain rate in the fluid and $\mathsfbi{I}$ the $3 \times 3$ identity matrix. The single dot denotes contraction of the last index of the preceding quantity with the first index of the following quantity, as scalar products of vectors and matrix-vector products. 

Due to the linearity of the Stokes equations, it exists a linear relation between on the one hand the velocities $\boldsymbol{U}_b$ and angular velocities $\boldsymbol{\Omega}_b$ of each particle, and, on the other hand, the hydrodynamic forces $\boldsymbol{F}^\mathrm{hydro}_b$ and torques $\boldsymbol{T}^\mathrm{hydro}_b$ that the fluid exerts on each particle $b = 1, 2$. 
To shorten notation for the subsequent calculation, we introduce each a generalised velocity and force vector,
\begin{equation}
\boldsymbol{\mathcal{U}} = (\boldsymbol{U}_1, \boldsymbol{U}_2, \boldsymbol{\Omega}_1, \boldsymbol{\Omega}_2), \ \ \boldsymbol{\mathcal{F}}^\mathrm{hydro} = (\boldsymbol{F}^\mathrm{hydro}_1, \boldsymbol{F}^\mathrm{hydro}_2, \boldsymbol{T}^\mathrm{hydro}_1, \boldsymbol{T}^\mathrm{hydro}_2). 
\label{eq:defGeneralizedUF}
\end{equation}
The components of both vectors will be referred to by upper indices $P$ or $Q$ which adopt the values $t$ (translation) or $r$ (rotation), and a lower index $b$ or $c$ denoting the particle number (1 or 2), e.g. $\boldsymbol{\mathcal{U}}^r_1 = \boldsymbol{\Omega}_1$. 
The resistance matrix with components $\mathsfbi{R}^{PQ}_{bc}$ then is defined by 
\begin{equation}
(\boldsymbol{\mathcal{F}}^\mathrm{hydro})^P_b =: \sum_{Q, c}\mathsfbi{R}^{PQ}_{bc} \boldsymbol{\cdot} \boldsymbol{\mathcal{U}}^Q_c. 
\label{eq:defResistanceMatrix}
\end{equation}
Since in the regime of low Reynolds numbers the particle inertia and thus acceleration forces are negligible compared to friction, the hydrodynamic forces on each particle have to equal the negative external forces applied, $\boldsymbol{F}_b = - \boldsymbol{F}^\mathrm{hydro}_b$, and similar for torques $\boldsymbol{T}_b = - \boldsymbol{T}^\mathrm{hydro}_b$. This implies 
\begin{equation}
\boldsymbol{\mathcal{F}} = - \boldsymbol{\mathcal{F}}^\mathrm{hydro},
\label{eq:externalForces}
\end{equation}
with $\boldsymbol{\mathcal{F}} = (\boldsymbol{F}_1, \boldsymbol{F}_2, \boldsymbol{T}_1, \boldsymbol{T}_2)$. 
The mobility matrix with components $\mathsfbilow{\mu}^{PQ}_{bc}$ is then defined as 
\begin{equation}
\boldsymbol{\mathcal{U}}^P_b =: \sum_{Q, c} \mathsfbilow{\mu}^{PQ}_{bc} \boldsymbol{\cdot} \boldsymbol{\mathcal{F}}^Q_c, 
\label{eq:defMobilityMatrix}
\end{equation}
and hence $\mathsfbilow{\mu} = - \mathsfbi{R}^{-1}$. 

For the case of constant fluid viscosity $\eta^{(0)}$, the mobility matrix is given up to order $s^{-2}$ by \citep[pp. 248-256]{Dhont1996}
\begin{equation}
\mathsfbilow{\mu}^{(0)} = \frac{1}{6 \pi a \eta^{(0)}} 
\begin{pmatrix}
\mathsfbi{I} & \mathsfbi{T} & \mathsfbi{0} & \mathsfbi{V} \\
\mathsfbi{T} & \mathsfbi{I} & \mathsfbi{-V} & \mathsfbi{0} \\
\mathsfbi{0} & \mathsfbi{V} & \frac{3}{4 a^2} \mathsfbi{I} & \mathsfbi{0} \\
\mathsfbi{-V} & \mathsfbi{0} & \mathsfbi{0} &  \frac{3}{4 a^2} \mathsfbi{I}
\end{pmatrix}, 
\end{equation}
with
\begin{equation}
\mathsfbi{T} := \frac{3 a}{4 s} \left( \mathsfbi{I} + \frac{\boldsymbol{s} \otimes \boldsymbol{s}}{s^2} \right),
\end{equation}
\begin{equation}
\mathsfbi{V} = \frac{3 a}{4 s^3} (\boldsymbol{s} \times),
\end{equation}
and the tensor product denoted by $\otimes$. Furthermore, $\boldsymbol{s} := \boldsymbol{r}_2 - \boldsymbol{r}_1$ is the separation vector between the positions of the two particles and $\mathsfbi{0}$ the $3 \times 3$ matrix with all entries zero. The tensor $(\boldsymbol{s} \times)$ is defined as $(\boldsymbol{s} \times ) := \mathsfbi{\epsilon}_{ikj} s_k \, \boldsymbol{e}_i \otimes \boldsymbol{e}_j$, with $\boldsymbol{e}_i$ the unit vector in the direction associated to the spatial index $i$.

\section{Derivation of the correction term to the mobility matrix}
\label{ch:perturbationAnalysis}
We now make use of the Lorentz reciprocal theorem to derive the first order correction in $\epsilon$ to the mobility matrix of a two-particle system, with particles at positions $\boldsymbol{r}_1$ and $\boldsymbol{r}_2$. Specifically, we consider two systems of which the first system has constant viscosity $\eta^{(0)}$ everywhere in the fluid and the second one is described by the perturbed viscosity profile $\eta(\boldsymbol{r})$ \citep{Oppenheimer2016}. Quantities in the first system will carry a superscript $^{(0)}$ associating them with constant viscosity, whereas quantities corresponding to the second system will carry no superscript in brackets. In both systems, the particles move and translate with velocities $\boldsymbol{\mathcal{U}}^{(0)}$ and $\boldsymbol{\mathcal{U}}$, respectively, inducing velocity, pressure and stress fields $\{\boldsymbol{u}^{(0)} (\boldsymbol{r}), p^{(0)} (\boldsymbol{r}), \mathsfbilow{\sigma}^{(0)} (\boldsymbol{r})\}$ and $\{\boldsymbol{u} (\boldsymbol{r}), p (\boldsymbol{r}), \mathsfbilow{\sigma} (\boldsymbol{r})\}$, respectively. To shorten notation, we will omit the position-dependence in the following and only use explicit notation for the sake of clarity when necessary. 
Following the derivation of the standard Lorentz reciprocal theorem (Appendix \ref{sec:app_1}, and \citet{Oppenheimer2016}), we arrive at the relation
\begin{equation}
\sum_{P, b} \boldsymbol{\mathcal{F}}^P_b \boldsymbol{\cdot} \boldsymbol{\mathcal{U}}^{(0) P}_b - \sum_{P, b} \boldsymbol{\mathcal{F}}^{(0) P}_b \boldsymbol{\cdot} \boldsymbol{\mathcal{U}}^P_b = 2 \int_V \left[ \eta(\boldsymbol{r}) - \eta^{(0)} \right] \mathsfbi{E} : \mathsfbi{E}^{(0)} dV.
\label{eq:intermediateStep}
\end{equation}
Here, $V$ denotes the fluid volume and $\mathsfbi{E} : \mathsfbi{E}^{(0)}$ the contraction of last two indices of $\mathsfbi{E}$ with the first two indices of $\mathsfbi{E}^{(0)}$. 
Naturally, while for $\eta(\boldsymbol{r}) = \eta^{(0)}$ one obtains the standard Lorentz reciprocal theorem, we here obtain a relation between the generalised velocities and forces and a correction term involving the viscosity perturbation and the strain rate in the fluid. 

For given particle positions, the fluid velocity at each point is a linear expression in the forces and torques acting on the particles. Therefore, we can express the strain rates $\mathsfbi{E}$ and $\mathsfbi{E}^{(0)}$ as products of normalised strain rates $\mathsfbi{B}$ and $\mathsfbi{B}^{(0)}$ with respect to the forces and torques acting on the particles, and the generalised forces: 
\begin{equation}
\mathsfbi{E} =: \sum_{P, b} \mathsfbi{B}^P_b \boldsymbol{\cdot} \boldsymbol{\mathcal{F}}^P_b, \ \ \mathsfbi{E}^{(0)} =: \sum_{P, b} \mathsfbi{B}^{(0) P}_b \boldsymbol{\cdot} \boldsymbol{\mathcal{F}}^{(0) P}_b. 
\label{eq:expansionStrainRates}
\end{equation}
Inserting this expansion together with Eq. \eqref{eq:defMobilityMatrix} and the analogous equation for the constant viscosity case into Eq. \eqref{eq:intermediateStep}, we obtain
\begin{eqnarray}
\sum_{P, Q, b, c} && \left( \boldsymbol{\mathcal{F}}^{P}_b \boldsymbol{\cdot} \mathsfbilow{\mu}^{(0) PQ}_{bc}  \boldsymbol{\cdot} \boldsymbol{\mathcal{F}}^{(0) Q}_c - \boldsymbol{\mathcal{F}}^{(0) P}_b \boldsymbol{\cdot} \mathsfbilow{\mu}^{PQ}_{bc} \boldsymbol{\cdot} \boldsymbol{\mathcal{F}}^{Q}_c \right)  \nonumber \\ = 2 &&  \int_V \left[ \eta(\boldsymbol{r}) - \eta^{(0)} \right] \sum_{P, Q, b, c} \left(\mathsfbi{B}^P_b \boldsymbol{\cdot} \boldsymbol{\mathcal{F}}^P_b \right) : \left(\mathsfbi{B}^{(0) Q}_c \boldsymbol{\cdot} \boldsymbol{\mathcal{F}}^{(0) Q}_c \right) dV. \label{eq:generalizedLRT}
\end{eqnarray}

Both sides are linear in $\boldsymbol{\mathcal{F}}$ and $\boldsymbol{\mathcal{F}}^{(0)}$, which are undetermined variables up to here. Hence, we obtain by suitably renaming the summation indices, using the symmetry of the mobility matrix \citep{Jeffrey1984} and factoring out $\boldsymbol{\mathcal{F}}$ and $\boldsymbol{\mathcal{F}}^{(0)}$ (Appendix \ref{sec:app_2}),
\begin{equation}
\mathsfbilow{\mu}^{PQ}_{bc} = \mathsfbilow{\mu}^{(0) PQ}_{bc} - 2 \int_V \left[ \eta(\boldsymbol{r}) - \eta^{(0)} \right] \mathsfbi{B}^P_b \doublecontract \mathsfbi{B}^{(0) Q}_c dV. 
\label{eq:correctionMobilityMatrix}
\end{equation}
Here, we followed the notation used in \citet{Oppenheimer2016}, defining 
\begin{equation}
\mathsfbi{B}^P_b \doublecontract \mathsfbi{B}^{(0) Q}_c := \sum_{k,l} \left(\mathsfbi{B}^P_b  \right)_{kli} \left( \mathsfbi{B}^{(0) Q}_c  \right)_{klj} \boldsymbol{e}_i \boldsymbol{e}_j.
\end{equation}
Realizing that $\eta(\boldsymbol{r}) - \eta^{(0)} = \epsilon \eta^{(1)} (\boldsymbol{r})$ and that we can expand $\mathsfbi{B} = \mathsfbi{B}^{(0)} + \textit{O}(\epsilon^1)$, we find that the $\epsilon^1$-correction to the mobility matrix only depends on the normalised strain rate in a constant-viscosity fluid, $\mathsfbi{B}^{(0) P}_b$, instead of $\mathsfbi{B}^P_b$. Consequently, we obtain for the $\epsilon^1$-correction to the mobility matrix
\begin{equation}
\mathsfbilow{\mu}^{(1) PQ}_{bc} = - 2 \int_V \eta^{(1)} (\boldsymbol{r}) \, \mathsfbi{B}^{(0) P}_b \doublecontract \mathsfbi{B}^{(0) Q}_c dV,
\label{eq:correctionMobilityGeneral}
\end{equation} 
which holds irrespective of the actual form of the gradient and the number of particles in the system. 
Note that in \citet{Oppenheimer2016}, the strain rates have been expanded with respect to the generalised particle velocities instead of the generalised forces. This yields a similar result for the first order correction to the resistance matrix rather than for the mobility matrix. 

While the normalised strain rate for a single particle can be directly obtained from the flow field associated with translation and rotation of a single particle, in the case of two or more particles one has to also account for the disturbance flows that other particles induce in response to the flow field produced by a first particle. Depending on whether the expansion of the strain rates is done with respect to generalised forces or velocities, these other particles must be assumed to be force- and torque-free, or to have zero (angular) velocity in the calculation of the normalised strain rates, respectively. 
By using a reflection method, one can account for all those disturbance flows ordered by ascending powers of the inverse particle separation \citep[Chapter 8]{KimKarilla1991}. 

At this point, expanding the strain rates with respect to the generalised forces simplifies the subsequent calculations drastically in comparison to an expansion with respect to the generalised velocities. The reason for this is that the disturbance flow field induced by a force- and torque-free particle scales to leading order with the local gradient of the flow field it is immersed in, namely the local strain rate of the flow, whereas a particle with zero velocity induces, in general, a disturbance flow proportional to the local flow field \citep[p. 189]{KimKarilla1991}. This is clear also intuitively: a force-free particle translates and rotates according to the surrounding flow field, whereas a particle with prescribed zero velocity must resist any external flow, inducing a stronger disturbance flow. 

A spherical particle located at the origin and subject to a force in a constant viscosity fluid produces a flow field given by $\mathsfbi{T}^\mathrm{trans} (\boldsymbol{r}) /(6 \pi \eta^{(0)} a)$ contracted with the external force, with \citep[p. 246]{Dhont1996}
\begin{equation}
    \mathsfbi{T}^\mathrm{trans} (\boldsymbol{r}) = \left( \frac{3a}{4r} \left(\mathsfbi{I} + \frac{\boldsymbol{r} \otimes \boldsymbol{r}}{r^2}  \right) + \frac{a^3}{4r^3} \left( \mathsfbi{I} - 3 \frac{\boldsymbol{r} \otimes \boldsymbol{r}}{r^2} \right)\right),
\end{equation}
the Rotne-Prager tensor and $r = |\boldsymbol{r}|$. 
Similarly, the flow field induced by a particle subject to a torque is given by $\mathsfbi{T}^\mathrm{rot} (\boldsymbol{r})/(8 \pi \eta^{(0)} a^3)$ contracted with the applied torque, with \citep[p. 248]{Dhont1996}
\begin{equation}
    \mathsfbi{T}^\mathrm{rot} (\boldsymbol{r}) = - \left(\frac{a}{r} \right)^3 (\boldsymbol{r} \times), 
\end{equation}
the rotlet. 
We then define the normalised single particle strain rates for translation and rotation as the symmetrized gradient of the corresponding flow field, where we use explicit index notation for the sake of clarity, 
\begin{equation}
     (\mathsfbi{P}^{(0) t})_{kli} (\boldsymbol{r}) :=  \frac{1}{6 \pi \eta^{(0)} a} \frac{1}{2} [\del_k  (\mathsfbi{T}^\mathrm{trans})_{li} (\boldsymbol{r})   + \del_l (\mathsfbi{T}^\mathrm{trans})_{ki} (\boldsymbol{r})],
\label{eq:Pt}
\end{equation}
\begin{equation}
    (\mathsfbi{P}^{(0) r})_{kli} (\boldsymbol{r}) :=  \frac{1}{8 \pi \eta^{(0)} a^3}\frac{1}{2} [\del_k  (\mathsfbi{T}^\mathrm{rot})_{li} (\boldsymbol{r})   + \del_l (\mathsfbi{T}^\mathrm{rot})_{ki} (\boldsymbol{r})].
\label{eq:Pr}
\end{equation}

The normalised strain rates in the two particle system, $\mathsfbi{B}^{(0) P}_b$, can then be expanded by means of the reflection method (Appendix \ref{sec:app_3}) in terms of the single particle strain rates as
\begin{equation}
    \mathsfbi{B}^{(0) t}_b \equiv \mathsfbi{B}^{(0) t}_b (\boldsymbol{r}) = \mathsfbi{P}^{(0) t} (\boldsymbol{r} - \boldsymbol{r}_b) + \mathsfbi{P}^{(0) s} (\boldsymbol{r} - \boldsymbol{r}_c) : \mathsfbi{P}^{(0) t} (\boldsymbol{r}_c - \boldsymbol{r}_b) + \textit{O}(s^{-3}),
\label{eq:expansionTranslationalNSR}
\end{equation}
and 
\begin{equation}
    \mathsfbi{B}^{(0) r}_b \equiv \mathsfbi{B}^{(0) r}_b (\boldsymbol{r}) = \mathsfbi{P}^{(0) r} (\boldsymbol{r} - \boldsymbol{r}_b) + \textit{O}(s^{-3}). 
\label{eq:expansionRotationalNSR}
\end{equation}
Here, $\mathsfbi{P}^{(0) s}$ denotes the normalised strain rate associated with the disturbance flow field that a spherical particle immersed in a local strain rate induces (see Appendix \ref{sec:app_3} for details). 

We have worked out the reflection scheme up to $\textit{O}(s^{-2})$, which will suffice to calculate the correction to the mobility matrix up to order $\textit{O}(s^{-1})$ in a linear viscosity gradient and up to order $\textit{O}(s^{-2})$ for particles with a temperature contrast to the fluid. 
Inserting the expansions \eqref{eq:expansionTranslationalNSR} and \eqref{eq:expansionRotationalNSR} into Eq. \eqref{eq:correctionMobilityGeneral}, we see
that the task of calculating the integrals \eqref{eq:correctionMobilityGeneral} reduces to the calculation of integrals of the form 
\begin{equation}
\mathsfbi{Q}^{(1) PQ}_{bc} := - 2 \int_V \eta^{(1)}(\boldsymbol{r}) \, \mathsfbi{P}^{(0) P} (\boldsymbol{r} - \boldsymbol{r}_b) \doublecontract \mathsfbi{P}^{(0) Q} (\boldsymbol{r} - \boldsymbol{r}_c) \, dV,
\label{eq:defQ}
\end{equation}
for $P, Q = t, r, s$, as discussed in the following sections.

\section{Particles in a large linear viscosity gradient}
\label{ch:linearGradient}
\begin{figure}
\begin{center}
\includegraphics[scale=0.5]{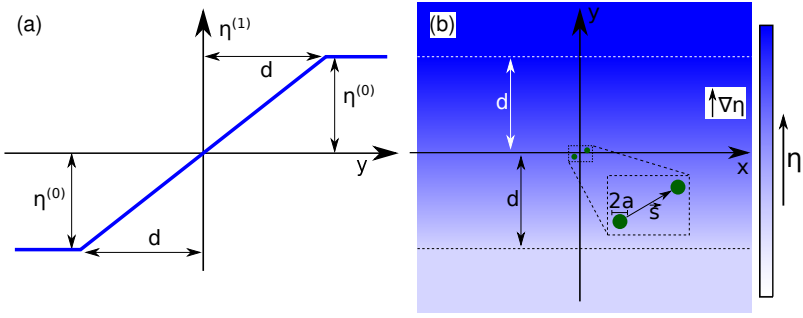}
\caption{Sketch of the prototypical interface-like viscosity gradient with the two interacting particles close to central plane of the gradient. (a) Schematics of the position-dependent viscosity. (b) Sketch of the fluid with darker blue denoting regions of higher viscosity and brighter blue denoting areas of lower viscosity.}
\label{fig:sketchViscGradInterface}
\end{center}
\end{figure}
We first consider the mobility matrix of two particles within a large linear viscosity gradient, i.e. with an extension $d$ larger by orders of magnitude than the distance $s$ between the two particles. In particular, we assume that the viscosity profile is given by 
\begin{equation}
\eta (\boldsymbol{r}) = \eta_0 + \epsilon \boldsymbol{H}^{(1)} \boldsymbol{\cdot} \boldsymbol{r}.
\end{equation}
for all $|\boldsymbol{r}| < d$, with $\boldsymbol{H}^{(1)}$ the vectorial gradient and $\eta_0$ a constant viscosity background. We introduce a second dimensionless parameter $\kappa := a/d$ as the ratio of bead radius and size of the viscosity gradient.

A prototypical interface-like linear viscosity gradient is, without restriction of generality along the $y$-direction, given by 
\begin{equation}
\eta (\boldsymbol{r}) = \eta_0 + 
\begin{cases}
- \epsilon \eta_0, &y < -d \\
(0, \epsilon \frac{\eta_0}{d}, 0) \boldsymbol{\cdot} \boldsymbol{r}, &-d \leq y \leq d\\
\epsilon \eta_0, &d < y,
\end{cases}
\label{eq:interfaceLikeGradient}
\end{equation}
with $\boldsymbol{r} = (x, y, z)$.
It describes a viscosity perturbation which is linear in an infinite slab of diameter $2d$ and is constant in the two-half spaces separated by the slab with a continuous transition (Fig. \ref{fig:sketchViscGradInterface}), avoiding negative viscosity. The perturbation parameter $\epsilon$ characterises the ratio of the maximal viscosity perturbation compared to the background viscosity. Such viscosity gradients can be expected to be found e.g. at the interface of two immiscible fluids of different viscosities.

For linear viscosity gradients, we will generally choose the reference viscosity $\eta^{(0)}$, which defines the $\epsilon^1$ viscosity perturbation as $\eta^{(1)} (\boldsymbol{r}) := (\eta(\boldsymbol{r}) - \eta^{(0)})/\epsilon$, to be the viscosity at some point close to the interacting particles. In comparison to a fixed reference viscosity independent of the actual positions of the particles, this choice is the more natural one as it minimises the viscosity perturbation close to the interacting particles and yields the most accurate results in the subsequent calculations. 
For this reason, $\eta_0$ and $\eta^{(0)}$ should be treated as different, since the latter varies with the particle positioning in the gradient while the former does not. 

We now calculate the mobility matrix at the first order $\epsilon^1 \kappa^1$, and include all terms up to $\textit{O}(s^{-1})$, i.e. focus on the far field interactions between the particles. 
We firstly consider the simpler case of an anti-symmetric viscosity gradient, like e.g. the interface-like gradient, under the assumption that both particles have distances from the centre of symmetry, or plane of symmetry in the case of the interface-like gradient, which are by orders of magnitude smaller than $d$, i.e. $|\boldsymbol{r}_b| \ll d$ (Section \ref{ch:symmetricLinearGradient}). 
Secondly, we will show that if either the particles are placed asymmetrically in an anti-symmetric gradient or if the gradient lacks this symmetry itself, an additional non-negligible contribution to the mobility matrix arises (Section \ref{ch:asymmetricLinearGradient}).

\subsection{Particles in an anti-symmetric viscosity gradient}
\label{ch:symmetricLinearGradient}
In this section, we assume anti-symmetry of the viscosity perturbation under $\boldsymbol{r} \to -\boldsymbol{r}$, i.e. $\eta^{(1)} (- \boldsymbol{r}) = - \eta^{(1)} (\boldsymbol{r})$, and both particles being close to the centre of symmetry, which we assume to be the origin of the coordinate system, $|\boldsymbol{r}_i| \ll d$. For the interface-like viscosity gradient, this is the case if the particles are positioned close to the plane characterised by $y = 0$ and thus we can in this case assume for the reference viscosity $\eta^{(0)} = \eta_0$. 
It can be shown that an infinite linear viscosity gradient, although inevitably associated with negative viscosity in major regions of the fluid, also yields mathematically the same result for the correction to the mobility matrix (Appendix \ref{sec:app_4.1}) up to order $\epsilon^1 \kappa^1$ as the anti-symmetric gradient, which allows to compare our results with previous works considering infinite gradients \citep{Datt2019a, Laumann2019}.

We defer the explicit calculation of the $\mathsfbi{Q}^{(1)}$-terms, which can be done with the help of a computer algebra system \citep{Mathematica2020}, to Appendix \ref{sec:app_4}. The results are presented in Eqs. \eqref{eq:QTermsLinGradSelfMob} and \eqref{eq:QTermLinGradInteraction}. We, nonetheless use these results to calculate the mobility matrix $\mathsfbilow{\mu}^{(1) PQ}_{bc}$ by inserting Eqs. \eqref{eq:expansionTranslationalNSR} and \eqref{eq:expansionRotationalNSR} into Eq. \eqref{eq:correctionMobilityGeneral} and obtain, after applying the reflection method, 
\begin{align}
&\mathsfbilow{\mu}^{(1) tt}_{bc} = \mathsfbi{Q}^{(1)tt}_{bc} + \sum_{e \neq c} \mathsfbi{Q}^{(1)ts}_{be} : \mathsfbi{P}^{(0) t} (\boldsymbol{r}_e - \boldsymbol{r}_c) + \sum_{d \neq b} \mathsfbi{P}^{(0) t} (\boldsymbol{r}_d - \boldsymbol{r}_b) \doublecontract \mathsfbi{Q}^{(1)st}_{dc} + \textit{O}(s^{-3}), \nonumber \\
&\mathsfbilow{\mu}^{(1) tr}_{bc} = \mathsfbi{Q}^{(1)tr}_{bc} + \sum_{d \neq b} \mathsfbi{P}^{(0) t} (\boldsymbol{r}_d - \boldsymbol{r}_b) \doublecontract \mathsfbi{Q}^{(1)sr}_{dc} + \textit{O}(s^{-3}), \label{eq:muInTermsOfQ} \\
&\mathsfbilow{\mu}^{(1) rr}_{bc} = \mathsfbi{Q}^{(1)rr}_{bc} + \textit{O}(s^{-3}), \nonumber
\end{align}
with $b, c, d$ and $e$ independent particle indices. We use the sum notation in Eqs. \eqref{eq:muInTermsOfQ} to cover both the cases $b = c$ and $b \neq c$. 
It can, however, be shown (Appendix \ref{sec:app_4.4}), that Eqs. \eqref{eq:muInTermsOfQ} reduce to 
\begin{equation}
\mathsfbilow{\mu}^{(1) PQ}_{bc} = \mathsfbi{Q}^{(1) PQ}_{bc} + \textit{O}(s^{-2}), 
\label{eq:simpleMuLinearGradient}
\end{equation}
for the linear viscosity gradient, i.e. all other summands in Eqs. \eqref{eq:muInTermsOfQ} scale at most as $\textit{O}(s^{-2})$ and can be neglected. 

Including all terms up to $\textit{O}(s^{-1})$, the $\epsilon^1 \kappa^1$ correction to the mobility matrix then is given by
\begin{eqnarray}
&&\mathsfbilow{\mu}^{(1)} = \frac{1}{6 \pi a (\eta^{(0)})^2} \times \label{eq:correctionMobility} \\ \nonumber
&& \begin{pmatrix}
- \eta^{(1)} (\boldsymbol{r}_1) \mathsfbi{I} & - \eta^{(1)} \left( \frac{\boldsymbol{r}_1 + \boldsymbol{r}_2}{2} \right) \mathsfbi{T} + \mathsfbi{W} & \frac{1}{4} (\bnabla \eta^{(1)} \times) & - \eta^{(1)} \left( \frac{\boldsymbol{r}_1 + \boldsymbol{r}_2}{2} \right) \mathsfbi{V} + \mathsfbi{Y} \\
- \eta^{(1)} \left( \frac{\boldsymbol{r}_1 + \boldsymbol{r}_2}{2} \right) \mathsfbi{T} - \mathsfbi{W} & -\eta^{(1)} (\boldsymbol{r}_2) \mathsfbi{I} &  \eta^{(1)} \left( \frac{\boldsymbol{r}_1 + \boldsymbol{r}_2}{2} \right) \mathsfbi{V} + \mathsfbi{Y}  & \frac{1}{4} (\bnabla \eta^{(1)} \times) \\
-\frac{1}{4} (\bnabla \eta^{(1)} \times) & - \eta^{(1)} \left( \frac{\boldsymbol{r}_1 + \boldsymbol{r}_2}{2} \right) \mathsfbi{V} + \mathsfbi{Y}^T & -\frac{3}{4 a^2} \eta^{(1)} (\boldsymbol{r}_1) \mathsfbi{I} & \mathsfbi{0} \\
 \eta^{(1)} \left( \frac{\boldsymbol{r}_1 + \boldsymbol{r}_2}{2} \right) \mathsfbi{V} + \mathsfbi{Y}^T & -\frac{1}{4} (\bnabla \eta^{(1)} \times) & \mathsfbi{0}  &  -\frac{3}{4 a^2} \eta^{(1)} (\boldsymbol{r}_2) \mathsfbi{I}
\end{pmatrix},
\end{eqnarray}
with
\begin{equation}
\mathsfbi{W} := \frac{3 a}{8 s} \left( \bnabla \eta^{(1)} \otimes \boldsymbol{s} - \boldsymbol{s} \otimes \bnabla \eta^{(1)} \right),
\end{equation}
and
\begin{equation}
\mathsfbi{Y} := \frac{3 a}{8 s^3} \boldsymbol{s} \otimes (\boldsymbol{s} \times \bnabla \eta^{(1)}).
\end{equation}
The corrections to the translational and rotational self-mobilities of both particles, i. e. the diagonal elements of Eq. \eqref{eq:correctionMobility}, are proportional to the negative local value of the viscosity perturbation, associating higher viscosity with lower bead self-mobilities as expected. In fact, the corrections to the self-mobilities correspond to the first order term in a Taylor expansion of the constant-viscosity expressions with the local viscosity $\eta (\boldsymbol{r}_b) = \eta^{(0)} + \epsilon \eta^{(1)} (\boldsymbol{r}_b)$ inserted, around $\epsilon = 0$. 

In contrast to the case of constant viscosity, the viscosity gradient couples the translational and rotational modes of a single particle, i.e. a particle experiences rotation when it is centrally subject to a force. Similarly, translation will appear if the particle is subject to a torque. 
Assuming a force $\boldsymbol{F}$ acting on a particle, the resulting angular velocity is given from Eq. \eqref{eq:correctionMobility} as 
\begin{equation}
\boldsymbol{\Omega} = -\frac{1}{24 \pi a (\eta^{(0)})^2} \bnabla \eta \times \boldsymbol{F},
\end{equation}
which can be understood as an effect of the imbalance of the friction forces \citep{Datt2019a}. On the side of higher viscosity, the particle experiences higher drag while on the side of lower viscosity the effect is inverse. When a force perpendicular to the viscosity gradient acts on a particle, it will therefore induce a rotation as if the particle adhered stronger on the side of higher viscosity.
Similarly, a particle subject to a torque moves translationally when the torque has a component orthogonal to the viscosity gradient. Our findings here reproduce exactly the results presented in \citet{Datt2019a} for infinite viscosity gradients. 

In a more intricate fashion, the viscosity gradient alters the interaction between two different particles. From the $\mathsfbilow{\mu}^{(1) tt}_{21}$ term in Eq. \eqref{eq:correctionMobility}, we see that the $\epsilon^1 \kappa^1$ component of the interaction between two particles composes of two terms. The first is the $\eta^{(1)} ((\boldsymbol{r}_1 + \boldsymbol{r}_2)/2) \, \mathsfbi{T}$ term, which can be understood as an expansion of the constant viscosity interaction, similarly to the $\mathsfbilow{\mu}^{(1) tt}_{bb}$ and $\mathsfbilow{\mu}^{(1) rr}_{bb}$ terms. Here, the viscosity is evaluated at the geometric centre between both particles. The second term in the $\epsilon^1 \kappa^1$-component of the interaction is associated with $\mathsfbi{W}$. It gives rise to novel effects, in particular because it introduces a velocity component orthogonal to the force, when the force is parallel to the connection of both particles. This component is non-zero as soon as the viscosity gradient has a component orthogonal to the connection of both particles. 

For the purpose of illustration, we derive the $\epsilon^1 \kappa^1$-correction to the flow field produced by a particle subject to some force $\boldsymbol{F}_1$ within the viscosity gradient from the $\mathsfbilow{\mu}^{(1) tt}_{21}$ component in Eq. \eqref{eq:correctionMobility}, by treating the second sphere as a test particle.  
\begin{figure}
\begin{center}
\includegraphics[scale=0.8]{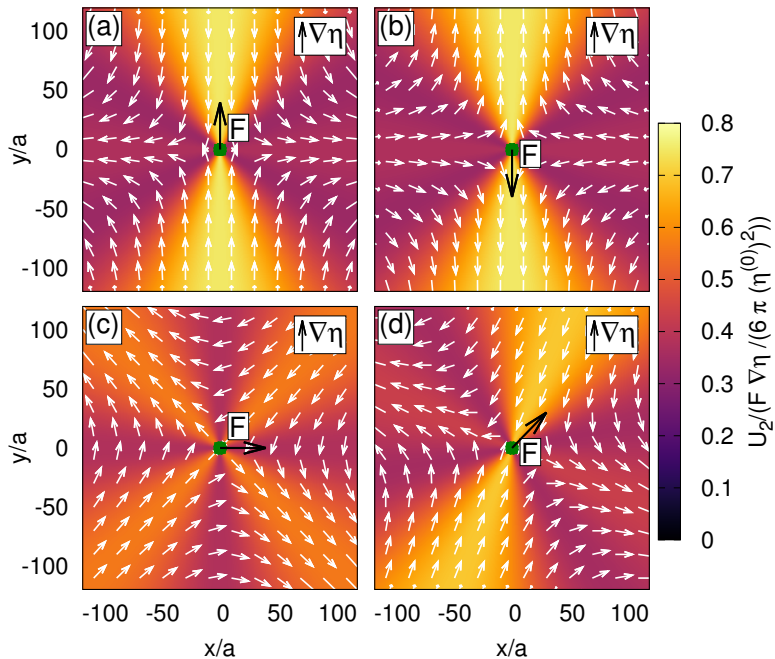}
\caption{The $\epsilon^1 \kappa^1$-contribution to the far field flow produced by a sphere subject to a force within the viscosity gradient for different directions of the force.}
\label{fig:flowField}
\end{center}
\end{figure}
We assume that the particle producing the flow is positioned at the origin of the coordinate system, $\boldsymbol{r}_1 = 0$, where we also assume the viscosity perturbation to vanish, i.e. $\eta^{(1)} (\boldsymbol{r}_1) = 0$. The $\epsilon^1$-correction to the flow field is then given as the velocity of the second particle located at $\boldsymbol{r}_2 = \boldsymbol{r}$, 
\begin{equation}
\boldsymbol{u}^{(1)}_F (\boldsymbol{r}) = \boldsymbol{U}_2 = \frac{1}{6 \pi a (\eta^{(0)})^2} \frac{3a}{8|\boldsymbol{r}|}\left[ - (\boldsymbol{r} \boldsymbol{\cdot} \bnabla \eta^{(1)}) \left( \mathsfbi{I} + \frac{\boldsymbol{r} \otimes \boldsymbol{r}}{|\boldsymbol{r}|^2} \right) - \left( \bnabla \eta^{(1)} \otimes \boldsymbol{r} - \boldsymbol{r} \otimes \bnabla \eta^{(1)} \right) \right] \boldsymbol{\cdot} \boldsymbol{F}_1,
\label{eq:flowField}
\end{equation}
where the first contribution comes from the $- \eta^{(1)} ((\boldsymbol{r}_1 + \boldsymbol{r}_2)/2) \, \mathsfbi{T}$ and the second contribution from the $\mathsfbi{W}$ term. 
This correction flow field is divergence-free and hence satisfies the incompressibility condition as expected. It is in agreement with the generalised Oseen tensor which was derived in \citet{Laumann2019}. 

When the force $\boldsymbol{F}$ is applied parallel to the viscosity gradient, it causes an $\epsilon^1 \kappa^1$-flow directed inwards along the direction of the viscosity gradient and a flow directed outwards in the direction orthogonal to the gradient (Fig. \ref{fig:flowField} (a)). When the force is acting in the opposite direction, i.e. down the viscosity gradient, also the correction flow field inverts (Fig. \ref{fig:flowField} (b)), as a direct consequence of the linearity of the problem. 
Qualitatively, this effect can be understood as follows: in the region above the particle, the fluid has higher viscosity and hence will react weaker to the force than a constant-viscosity fluid. Therefore, the correction flow field is oriented oppositely to the force here. In the region below the particle with decreased viscosity, the fluid reacts stronger and the correction flow is therefore parallel to the force. 

When the force is acting perpendicular to the viscosity gradient (Fig. \ref{fig:flowField} (c)), the correction flow again follows the direction of the force in the region of lower viscosity and opposes it in the region of higher viscosity. On the axis perpendicular to the viscosity gradient, a correction flow perpendicular to the force applied is found, such that again a pattern of inwards flow and outwards flow is observed.
When the force on the particle at the origin acts in some arbitrary direction (Fig. \ref{fig:flowField} (d)), the resulting flow field will be a linear superposition of the flow fields in Fig. \ref{fig:flowField} (a) and (c).

As an important consequence of Eq. \eqref{eq:correctionMobility}, the far field correction flow at position $\boldsymbol{r}$ depends only on the direction of $\boldsymbol{r}$ but scales constant in the absolute value of the distance, i.e. it does not decay. 
This contrasts with the constant-viscosity Oseen tensor, which decays with the first power of the inverse distance from the particle.
However, this correction flow field is only valid as long as the particle separation is, by orders of magnitude, smaller than $d$. Therefore, the correction flow, being proportional to $\epsilon \kappa$, always compares small to the Stokeslet flow field. 

The $\mathsfbilow{\mu}^{(1) rt}_{21}$ term in Eq. \eqref{eq:correctionMobility} corresponds to the angular velocity $\boldsymbol{\Omega}^{(1)}_F$ of particle 2 when particle 1 is subject to a force. With $\boldsymbol{r}_1 = 0$ and $\boldsymbol{r}_2 = \boldsymbol{r}$, we find
\begin{equation}
\boldsymbol{\Omega}^{(1)}_F(\boldsymbol{r}) = \boldsymbol{\Omega}_2 = \frac{1}{6 \pi a (\eta^{(0)})^2} \frac{3a}{8|\boldsymbol{r}|^3}\left[ (\boldsymbol{r} \boldsymbol{\cdot} \bnabla \eta) (\boldsymbol{r} \times) + (\boldsymbol{r} \times \bnabla \eta^{(1)}) \otimes \boldsymbol{r} \right] \boldsymbol{\cdot} \boldsymbol{F}_1,
\end{equation}
where the first summand results from $\eta^{(1)} ((\boldsymbol{r}_1 + \boldsymbol{r}_2)/2) \mathsfbi{V}$ and the second summand from $\mathsfbi{Y}^T$. 
A straight-forward calculation verifies that indeed $\boldsymbol{\Omega}_F (\boldsymbol{r}) = \bnabla \times \boldsymbol{u}^{(1)}_F (\boldsymbol{r})/2$, i.e. the particle rotation can consistently be also obtained from the flow field given in Eq. \eqref{eq:flowField} by invoking Faxen's second law \citep[p. 255]{Dhont1996}. 

From the $\mathsfbilow{\mu}^{(1) tr}_{21}$ term in the mobility matrix \eqref{eq:correctionMobility}, we extract the flow field that a particle subject to a torque induces in the viscosity gradient. 
Assuming that, similarly to before, the viscosity perturbation vanishes at the position of the first particle, the velocity of the second particle acting as a test particle is then given as
\begin{equation}
\boldsymbol{u}^{(1)}_T (\boldsymbol{r}) = \boldsymbol{U}_2 = \frac{1}{6 \pi a (\eta^{(0)})^2} \frac{3a}{8|\boldsymbol{r}|^3} \left[ (\bnabla \eta^{(1)} \boldsymbol{\cdot} \boldsymbol{r}) (\boldsymbol{r} \times \boldsymbol{T}_1) + \boldsymbol{r} \left( (\boldsymbol{r} \times \bnabla \eta^{(1)}) \boldsymbol{\cdot} \boldsymbol{T}_1 \right) \right].
\end{equation}
\begin{figure}
\begin{center}
\includegraphics[scale=0.4]{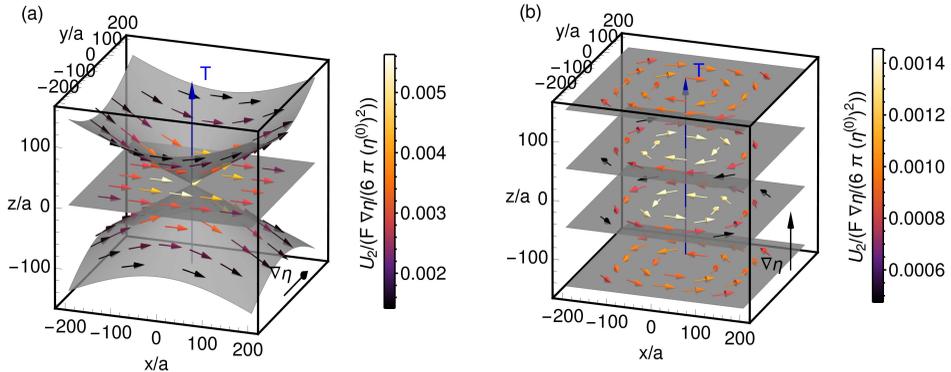}
\caption{Far field $\epsilon^1 \kappa^1$ correction flow induced by a particle subject to an externally applied torque along the direction indicated by the blue arrow with the direction of the viscosity gradient (a) orthogonal to it and (b) parallel to it. The grey surfaces are a guide to the eye and are chosen parallel to the flow field.}
\label{fig:correctionFlowFromTorque}
\end{center}
\end{figure}
If a torque acts orthogonal to the viscosity gradient (Fig. \ref{fig:correctionFlowFromTorque} (a)), a correction flow predominantly in the direction of $\bnabla \eta \times \boldsymbol{T}_1$ is induced. This result holds strictly at $z = 0$, while for $z \neq 0$ an additional component of the flow in $z$-direction is found. This finding is intuitive since the particle itself moves in the same direction as a result of the correction to its self-mobility. 
For a torque parallel to the viscosity gradient (Fig. \ref{fig:correctionFlowFromTorque} (b)), an additional rotating flow is observed which has the direction of rotation prescribed by $\boldsymbol{T}_1$ in the region of lower viscosity (here $z < 0$) and opposite direction in the region of increased viscosity (here $z > 0$). 
In the plane $z = 0$, the correction flow vanishes. Also this is intuitive, because the fluid opposes the rotation stronger in the region of higher viscosity, thus resulting in a correction flow with opposite direction of rotation, and the opposite effect in the lower-viscosity region. 

In both cases, the correction flow decays radially as $r^{-1}$, i.e. by one order slower than in the constant viscosity case where the rotlet decays as $r^{-2}$. 
Similarly to the flow field induced by a particle subject to a force, the correction flow field resulting from the linear viscosity gradient decays by one order slower than the constant-viscosity field does. 
A linear viscosity gradient thus induces long-range interactions between the particles, as long as the particle separation is well shorter than the size of the gradient. 

The results presented here for the interaction of particles in a linear viscosity gradient are also valid for more than two particles (Appendix \ref{sec:app_6}) as three-body effects scale higher in $s^{-1}$ and thus particle interaction is pairwise at the level of approximation chosen here.

\subsection{Asymmetric particle placement in the viscosity gradient}
\label{ch:asymmetricLinearGradient}
In the previous section, we have assumed that the viscosity perturbation is anti-symmetric and both particles are located close to the centre of symmetry or the central plane of the interface-like gradient when compared to the gradient size $d$. 
Since in a realistic setting the particle positions in the viscosity gradient are arbitrary, we here consider the interacting particles at varying positions within a finite-size gradient and calculate the effect of the viscosity gradient compared to the interaction in constant viscosity with the local viscosity as reference. Yet, we require that firstly the distance between both particles is still smaller than the gradient size, $s \ll d$, and that secondly the distance of each particle to the boundary of the linear gradient in any spatial direction is of the same order of magnitude as $d$, i.e. we assume that both particles do not come close to the edge of the linear gradient at length scales of $d$. 

Under these assumptions, it can be shown (Appendix \ref{sec:app_4.1}) that both the translational self-mobilities $\mathsfbilow{\mu}^{(1) tt}_{bb}$ as well as the translational interactions $\mathsfbilow{\mu}^{(1) tt}_{bc}$ with $b \neq c$, as obtained for an anti-symmetric viscosity gradient, have to be corrected by an additional term given by
\begin{equation}
\mathsfbi{Q}^{(1) tt}_\mathrm{asym} = - \frac{2}{(6 \pi a \, \eta^{(0)})^2} \int_V \left[ \eta^{(1)} (\boldsymbol{r}) - \tilde{\eta}^{(1)} (\boldsymbol{r}) \right] \frac{27 a^2}{8 |\boldsymbol{r} - \boldsymbol{r}_0|^6} (\boldsymbol{r} - \boldsymbol{r}_0) \otimes (\boldsymbol{r} - \boldsymbol{r}_0) \, dV,
\label{eq:QTermAsymmetry}
\end{equation}
with $\tilde{\eta}^{(1)} = \eta_0 + \boldsymbol{H}^{(1)} \boldsymbol{\cdot} \boldsymbol{r}$ the linear viscosity gradient extended to infinity and $\boldsymbol{r}_0$ a point close to the positions of both particles, i.e. $|\boldsymbol{r}_b - \boldsymbol{r}_0| \ll d, \ b = 1,2$. A similar correction term arises also if the domain of the viscosity gradient itself is not symmetric under point reflection. This correction term quantifies the error made by assuming an infinite viscosity gradient instead of a finite one in the calculation of the $\mathsfbi{Q}^{(1)}$ terms. All other $\mathsfbi{Q}^{(1)}$ terms are unchanged at this order of $\kappa$, such that the mobility matrix must be corrected by the additional term
\begin{equation}
\mathsfbilow{\mu}^{(1)}_\mathrm{asym} =
\begin{pmatrix}
\mathsfbi{Q}^{(1) tt}_\mathrm{asym} & \mathsfbi{Q}^{(1) tt}_\mathrm{asym} & \mathsfbi{0} & \mathsfbi{0} \\
\mathsfbi{Q}^{(1) tt}_\mathrm{asym} & \mathsfbi{Q}^{(1) tt}_\mathrm{asym} & \mathsfbi{0} & \mathsfbi{0} \\
\mathsfbi{0} & \mathsfbi{0} & \mathsfbi{0} & \mathsfbi{0} \\
\mathsfbi{0} & \mathsfbi{0} & \mathsfbi{0} & \mathsfbi{0} \\
\end{pmatrix}.
\end{equation}

This correction yields an additional velocity term for both particles 1 and 2, when the sum of the forces on both particles does not vanish: 
\begin{equation}
\boldsymbol{U}_1^\mathrm{asym} = \boldsymbol{U}_2^\mathrm{asym} = \mathsfbi{Q}^{(1) tt}_\mathrm{asym} \boldsymbol{\cdot} (\boldsymbol{F}_1 + \boldsymbol{F}_2),
\end{equation}
Notably, it is the sum of both forces which induces the additional velocity, such that a force-free particle 1 experiences the same additional velocity as particle 2, if particle 2 is subject to a force. 

\begin{figure}
\begin{center}
\includegraphics[scale=0.9]{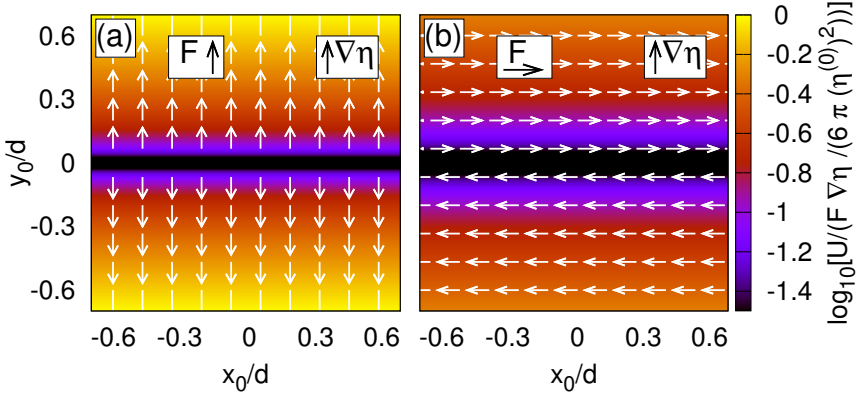}
\caption{Additional velocity $\boldsymbol{U} = \boldsymbol{U}_1^\mathrm{asym}$ due to asymmetric particle placement in dependence on the position of both particles in the interface-like viscosity gradient. Since the particles have relative distance $s \ll d$, they can be considered as located at the same position at scales of $d$. (a) The correction from asymmetry for a force $\boldsymbol{F} = \boldsymbol{F}_1 + \boldsymbol{F}_2$ applied along the $y$-axis, (b) the correction for a force applied along the $x$-axis.}
\label{fig:QTermAsymmetry}
\end{center}
\end{figure}

To illustrate this effect, we consider a particle in the interface-like gradient at position $y_0 = y'_0 d$, with $y'_0$ independent of $d$ and $\eta^{(0)}$ the reference viscosity at $y_0$. 
The correction term $\mathsfbi{Q}^{(1) tt}_\mathrm{asym}$ in dependence of $y_0$ is then calculated in Appendix \ref{sec:app_4.5} as
\begin{equation}
\mathsfbi{Q}^{(1) tt}_\mathrm{asym} = \frac{1}{6 \pi a (\eta^{(0)})^2} \frac{9 a |\bnabla \eta^{(1)}|}{16} \arctanh(y_0/d) \left( \mathsfbi{I} +  \hat{y} \otimes \hat{y} \right). 
\label{eq:correctionAsymmetry}
\end{equation}
The dependence of this correction term on the relative position in the viscosity gradient is illustrated in Fig. \ref{fig:QTermAsymmetry}, with the arrows indicating the direction and the color indicating the strength of $\boldsymbol{U}_1^\mathrm{asym} = \boldsymbol{U}_2^\mathrm{asym}$.

In the half-slab with higher viscosity ($y_0 > 0$ in Fig. \ref{fig:QTermAsymmetry}), $\boldsymbol{U}_1^\mathrm{asym}$ is parallel to the force, while in the other half of the slab it is directed oppositely to the force. Consistently with the results for anti-symmetric gradients, $\boldsymbol{U}_1^\mathrm{asym}$ vanishes in the central plane of the gradient. 
Intuitively, this result can be understood as follows: particles placed in the half-slab of higher viscosity have smaller distance to the edge of the linear gradient at higher viscosity than to the low-viscosity edge. Then, relative to the particle position, the fraction of the linear gradient with higher viscosity is smaller than the fraction with lower viscosity. In comparison to an anti-symmetric gradient, where those two fractions are the same, a correction along the force is expected and observed, since the anti-symmetric gradient overestimates the viscosity in parts of the fluid. For particles placed in the lower viscosity half-slab, the effect inverses. 

\begin{figure}
\begin{center}
\includegraphics[scale=0.9]{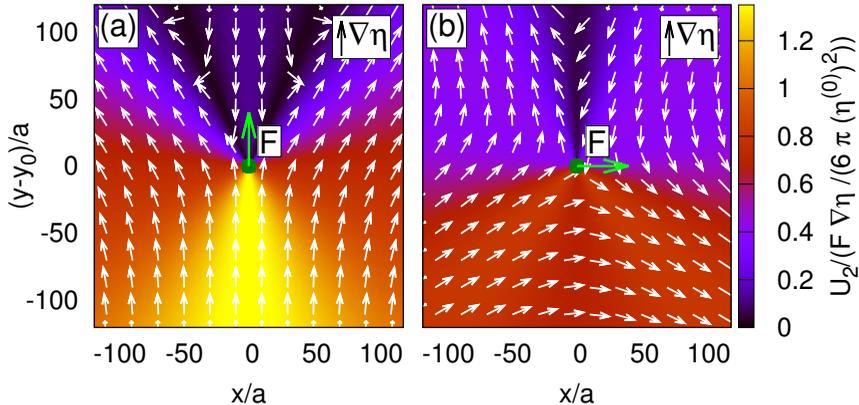}
\caption{The $\epsilon^1$ correction far field flow produced by a spherical particle subject to a force and placed asymmetrically in the interface-like viscosity gradient at $y_0 = 0.5 \, d$, where the local viscosity at the particle is taken as reference. In (a), the force acts along the $y$-direction, in (b) it acts along the $x$-direction.}
\label{fig:flowFieldAsymmetricallyPlacedParticle}
\end{center}
\end{figure}
For $|y_0| \lesssim d/2$, we have $\arctanh(y_0/d) \approx y_0/d = \textit{O}(1)$, and $\mathsfbi{Q}^{(1) tt}_\mathrm{corr}$ is of the same order of magnitude as the $\mathsfbi{W}$ term in Eq. \eqref{eq:correctionMobility}. 
The correction due to asymmetry therefore alters the $\epsilon^1$ correction flow as soon as the particles are placed away from the plane of symmetry of the interface-like gradient by a non-negligible fraction of $d$. In Fig. \ref{fig:flowFieldAsymmetricallyPlacedParticle}, the correction flow field produced by a particle located at $y_0 = 0.5 d$ with respect to the constant-viscosity flow field at the local viscosity as a reference is displayed. Comparison to the correction flow in a symmetric viscosity gradient (Fig. \ref{fig:flowField}) shows that an asymmetry in either the particle placement within the gradient or, similarly, an asymmetry in the gradient itself has a significant impact on the particle interactions. 
As a general rule, we find that the effect from asymmetry leads to a stronger correction flow field towards the central plane of the gradient and to a weaker correction field towards the edge of the gradient. The fact that the interaction is altered by an asymmetry in the viscosity gradient, although the gradient is locally unchanged, points out that the correction flows in a large linear viscosity gradient are a highly non-local effect. 

\section{Interaction of hot particles}
\label{ch:hotParticles}
As a second application of the presented method, we calculate the interaction of spherical particles with a homogeneous surface temperature that is different from the temperature of the surrounding fluid. The dependence of fluid viscosity on temperature can often be approximated using Andrade's equation, $\eta = A \exp(b/T)$ \citep{Andrade1930}, with $A$ and $b$ fluid-specific parameters. However, for small temperature contrasts, Taylor expanding Andrade's equation or one of its generalisations \citep{Gutmann1952} yields a linear relation between temperature and fluid viscosity, which we will use in the following. 
As a result, a single hot particle induces a viscosity perturbation decaying with the inverse distance from the particle centre, which in turn alters the self-mobility of the particle as shown in \citet{Oppenheimer2016}. However, to the best of our knowledge no results are available for the interaction of two hot particles, which we determine in the following. 
\begin{figure}
\begin{center}
\includegraphics[scale=0.5]{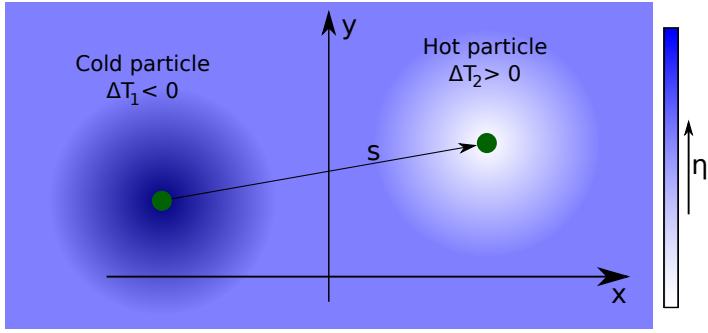}
\caption{Exemplary sketch of the viscosity perturbation in the presence of a hot and cold particle. The fluid viscosity is smaller in the neighbourhood of the hot particle and higher in the neighbourhood of the cold particle.}
\label{fig:sketchHotParticles}
\end{center}
\end{figure}

For a two-particle system with prescribed homogeneous surface temperatures, we firstly assume that a steady state is established by thermal conduction. Consequently, our model requires that the steady state establishes on faster time scales than the particle temperature changes. In practice, this is the case for beads with large heat capacities. The same would occur for objects subject to constant heat radiation if the susceptibility of the particles is different from that of the surrounding fluid. 
Secondly, our approach requires a small Peclet number (\textit{Pe}), which compares advective to thermal heat transport 
\begin{equation}
    \mathrm{\textit{Pe}} = \frac{U a}{\kappa} = \mathrm{\textit{Re}} \, \mathrm{\textit{Pr}},
\end{equation}
with $U$ the typical velocity in the fluid and $\kappa$ the thermal diffusivity. \textit{Re} and \textit{Pr} denote the Reynolds and Prandtl number, of which the latter depends only on properties of the fluid \citep[p. 596]{Leal2007}. 

The conduction-induced temperature field around a particle with homogeneous surface temperature in a quiescent fluid, which decays as $r^{-1}$, typically breaks down at a length scale of $\textit{O}(a / \mathrm{\textit{Pe}})$ \citep[p. 604]{Leal2007}. 
For the subsequent calculation we assume that $s \ll a/ \mathrm{\textit{Pe}}$, i.e. the distance between the two interacting particles is smaller than the length scale at which the temperature profile in the fluid becomes advection-dominated. Similarly to the single particle mobilities \citep{Oppenheimer2016}, the corresponding correction due to finite \textit{Pe} then scales asymptotically as $\textit{O}(\mathrm{\textit{Pe}}^2)$ and can be neglected in the limit of small Peclet numbers.
Indeed, for e.g. a micron-sized particle moving at a speed of microns per second in water at room temperature, the Reynolds number is $\textit{O}(10^{-6})$ and, with $\mathrm{\textit{Pr}} \sim \textit{O}(10)$ \citep[Appendix B]{Dincer2015}, the Peclet number is $\textit{O}(10^{-5})$, such that the above assumptions are indeed valid for e.g. colloidal particles \citep{Lu2013} separated by tens of radii. 

Under these assumptions, the temperature field in the fluid is described by the heat equation with prescribed homogeneous temperatures at the particle surfaces ($T_1$ and $T_2$ respectively) and at infinity ($T_0$) as boundary conditions, 
\begin{equation}
\nabla^2 T(\boldsymbol{r}) = 0, \ \  T(\boldsymbol{r}) = T_b \ \mathrm{for} \ \boldsymbol{r} \in S_b, \ \ T(\boldsymbol{r}) \to T_0 \ \mathrm{for} \ |\boldsymbol{r}| \to \infty, 
\label{eq:steadyStateHeatEquation}
\end{equation}
with $S_b$ the surface of particle $b$. 
For a system consisting of two or more particles, the temperature field can be approximated by a reflection scheme, where monopolar, dipolar, etc. single particle fields are superposed to match the boundary conditions up to successively increasing order in the inverse particle separation, similarly to the fluid velocity field. 
We obtain for the temperature field including all terms up to $\textit{O}(s^{-2})$, 
\begin{equation}
T(\boldsymbol{r}) = T_0 + C^m_1 \frac{a}{|\boldsymbol{r} - \boldsymbol{r}_1|} + \frac{a^2 \boldsymbol{C}^d_1 \boldsymbol{\cdot} (\boldsymbol{r} - \boldsymbol{r}_1)}{|\boldsymbol{r} - \boldsymbol{r}_1|^3} +  C^m_2 \frac{a}{|\boldsymbol{r} - \boldsymbol{r}_2|} + \frac{a^2 \boldsymbol{C}^d_2 \boldsymbol{\cdot} (\boldsymbol{r} - \boldsymbol{r}_2)}{|\boldsymbol{r} - \boldsymbol{r}_2|^3} + \textit{O}(s^{-3}),
\label{eq:temperatureField}
\end{equation}
with 
\begin{align}
C^m_1 = \Delta T_1 \left(1 + \frac{a^2}{s^2} \right) - \Delta T_2 \frac{a}{s}, \ \boldsymbol{C}^d_1 = - \Delta T_2 \frac{a^2 \boldsymbol{s}}{s^3}, \\
C^m_2 = \Delta T_2 \left( 1 + \frac{a^2}{s^2} \right) - \Delta T_1 \frac{a}{s}, \ \boldsymbol{C}^d_2 = \Delta T_1 \frac{a^2 \boldsymbol{s}}{s^3}.
\end{align}

Approximating the dependence of the viscosity on the temperature as linear, we find \begin{equation}
\eta \approx \eta^{(0)} + \left. \frac{\del \eta}{\del T} \right|_{T_0}  (T - T_0) =: \eta^{(0)} + \alpha \, (T - T_0),
\label{eq:viscosityOnTemperature}
\end{equation}
with $\alpha$ the proportionality coefficient. For particles with a temperature contrast relative to the fluid, it is natural to assume $\eta^{(0)}$ to be the viscosity at temperature $T_0$. 
Assuming that $\Delta T_1$ and $\Delta T_2$ are typically of the order of some $\Delta T$, we define the perturbation parameter $\epsilon := - \alpha \Delta T/\eta^{(0)}$ as the relative change in viscosity associated with $\Delta T$ compared to $\eta^{(0)}$ \citep{Oppenheimer2016}. 
Typically, $\alpha$ is negative, associating hot particles ($T > T_0$) with a decrease in viscosity and cold particles ($T < T_0$) with an increase, and thus $\epsilon$ is positive \citep{Gutmann1952}. 

Using Eq. \eqref{eq:viscosityOnTemperature} and the definition of $\epsilon$, the $\epsilon^1$ viscosity perturbation is given as 
\begin{equation}
    \eta^{(1)} = \eta^{(0)} \frac{T(\boldsymbol{r}) - T_0}{\Delta T}.
\end{equation}
By inserting the expression for the temperature field, Eq. \eqref{eq:temperatureField}, we obtain the explicit expansion of the viscosity field,
\begin{equation}
\eta^{(1)}(\boldsymbol{r}) = \eta^{(1) m}_1 \frac{a}{|\boldsymbol{r} - \boldsymbol{r}_1|} + \frac{a^2 \boldsymbol{\eta}^{(1) d}_1 \boldsymbol{\cdot} (\boldsymbol{r} - \boldsymbol{r}_1)}{|\boldsymbol{r} - \boldsymbol{r}_1|^3} +  \eta^{(1) m}_2 \frac{a}{|\boldsymbol{r} - \boldsymbol{r}_2|} + \frac{a^2 \boldsymbol{\eta}^{(1) d}_2 \boldsymbol{\cdot} (\boldsymbol{r} - \boldsymbol{r}_2)}{|\boldsymbol{r} - \boldsymbol{r}_2|^3} + \textit{O}(s^{-3}),
\label{eq:viscosityPerturbationHotPart}
\end{equation}
with $\eta^{(1) m}_b = -(\eta^{(0)} C^m_b)/\Delta T, \ \ \boldsymbol{\eta}^{(1) d}_b = -(\eta^{(0)} \boldsymbol{C}^d_b)/\Delta T, \ \forall b \in \{1, 2\}$. 

Using Eq. \eqref{eq:viscosityPerturbationHotPart}, we first calculate the respective $\mathsfbi{Q}^{(1)}$ terms (Appendix \ref{sec:app_5}) and from this compute the $\epsilon^1$ correction to the mobility matrix. One can show that, up to $\textit{O}(s^{-2})$, the corrections to the mobility matrix are given by the corresponding $\mathsfbi{Q}^{(1)}$ terms (Appendix \ref{sec:app_5.3}),
\begin{equation}
\mathsfbilow{\mu}^{(1)PQ}_{bc} = \mathsfbi{Q}^{(1)PQ}_{bc} + \textit{O}(s^{-3}).
\label{eq:muEqualsQHot}
\end{equation}
We then find the $\epsilon^1$-correction to the mobility matrix given by
\begin{eqnarray}
&&\mathsfbilow{\mu}^{(1)tt}_{bb} = \frac{1}{6 \pi a \eta^{(0)}} \left[ \left( \left(\frac{5}{12} - \frac{7 a^2}{12 s^2} \right) \frac{\Delta T_b}{\Delta T} + \left( \frac{7 a}{12 s} - \frac{9 a^2}{8 s^2} \right) \frac{\Delta T_c}{\Delta T} \right) \mathsfbi{I} + \frac{9 a^2}{8} \frac{\Delta T_c}{\Delta T} \frac{\boldsymbol{s} \otimes \boldsymbol{s}}{s^4} \right] + \textit{O}(s^{-3}), \nonumber \\
&&\mathsfbilow{\mu}^{(1)rr}_{bb} = \frac{1}{6 \pi a \eta^{(0)}} \left[ \left(\frac{9}{16 a^2} - \frac{3}{16 s^2} \right) \frac{\Delta T_b}{\Delta T} + \frac{3}{16 as} \frac{\Delta T_c}{\Delta T} \right] \mathsfbi{I} + \textit{O}(s^{-3}), \nonumber \\
&&\mathsfbilow{\mu}^{(1) tr}_{11} = \frac{1}{6 \pi a \eta^{(0)}} \left( -\frac{\Delta T_2}{\Delta T} \frac{a (\boldsymbol{s} \times)}{8 s^3} \right) + \textit{O}(s^{-3}) = -\mathsfbilow{\mu}^{(1) rt}_{11}, \label{eq:mobilityHotParticles}\\
&&\mathsfbilow{\mu}^{(1) tr}_{22} = \frac{1}{6 \pi a \eta^{(0)}} \left( \frac{\Delta T_1}{\Delta T} \frac{a (\boldsymbol{s} \times)}{8 s^3} \right) + \textit{O}(s^{-3}) = -\mathsfbilow{\mu}^{(1) rt}_{22}, \nonumber \\
&&\mathsfbilow{\mu}^{(1)tt}_{bc} = \frac{1}{6 \pi a \eta^{(0)}} \left( \frac{9 a^2}{8} \frac{\Delta T_b + \Delta T_c}{\Delta T} \frac{\boldsymbol{s} \otimes \boldsymbol{s}}{s^4} \right) + \textit{O}(s^{-3}), \ b \neq c, \nonumber \\
&&\mathsfbilow{\mu}^{(1)tr}_{bc} = \mathsfbilow{\mu}^{(1) rt}_{bc} = \mathsfbilow{\mu}^{(1) rr}_{bc} = \textit{O}(s^{-3}), \ b \neq c \nonumber.
\end{eqnarray}
The corrections to the translational and rotational self-mobilities of, say, particle 1 depend at order $s^0$ only on the particle's own temperature, which alters the self-mobility similarly as in the case when there is only one particle in the fluid \citep{Oppenheimer2016}. If the particle is hot, i.e. $T_1 > T_0$, the particles mobility increases, and decreases for $T_1 < T_0$.

In contrast to the case of a single particle, we find an additional term scaling as $s^{-1}$ in the first particle's self-mobilities, which is proportional to the temperature of the second particle at the distance $s$. This is, because the temperature of particle 2 induces a radially decaying temperature and viscosity field around it, which can be approximated to leading order as a constant near particle 1. In order to satisfy the boundary condition for the temperature field at the surface of particle 1, a reflected monopolar temperature field around particle 1 is induced. However, the reflected flow field decays as $r^{-1}$ with $r$ the distance from particle 1. Therefore, the effect due to the incident field is stronger than the effect of the reflected field, and the mobility of particle 1 indeed increases for a hot particle 2 and decreases if particle 2 has a lower temperature than the fluid at infinity. 

While for a single particle with homogeneous surface temperature translation and rotation do not couple \citep{Oppenheimer2016}, such coupling is observed at $\textit{O}(s^{-2})$ in the presence of a second particle. 
This effect on say particle 1 can be understood as the result of the gradient of the monopolar temperature field induced by particle 2. 
However, also here a dipolar temperature field reflected from particle 1 arises to satisfy the boundary conditions at the surface of particle 1 and weakens the effect of the incident temperature gradient. 

Considering the interaction components in Eq. \eqref{eq:mobilityHotParticles}, only the translational terms are non-vanishing at our level of approximation. 
Interestingly, the translational interaction between two particles increases proportional to the sum of $\Delta T_1 + \Delta T_2$, i.e. it is irrelevant whether the particle causing or experiencing the interaction has an increased temperature. This observation is in agreement with the fact that the mobility matrix is symmetric as a result of the Lorentz reciprocal theorem \citep{Jeffrey1984}, a result that holds independently of whether the viscosity is constant or not. 
Moreover, if one particle is hotter and the other is colder than the fluid by the same amount, both effects cancel and the interaction between both particles is up to order $\textit{O}(s^{-2})$ similar to the interaction of two neutral particles. 
\begin{figure}
\begin{center}
\includegraphics[scale=0.8]{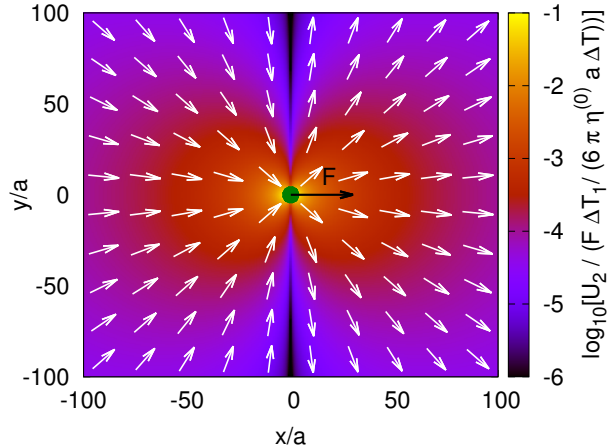}
\caption{The $\epsilon^1$ far field correction flow field produced by a hot particle with a force applied along the $x$-direction.}
\label{fig:correctionFlowHotParticle}
\end{center}
\end{figure}

Considering the second particle a neutral test particle, i.e. $\Delta T_2 = 0$, we find for the far field correction flow produced by particle 1 located at $\boldsymbol{r} = \boldsymbol{0}$ with increased temperature $\Delta T_1 > 0$ and subject to a force $\boldsymbol{F}_1$,
\begin{equation}
\boldsymbol{u}^{(1)} (\boldsymbol{r}) = \boldsymbol{U}_2 =  \frac{1}{6 \pi a \eta^{(0)}} \frac{9 a^2 \Delta T_1}{8 r^4 \Delta T} \boldsymbol{r} (\boldsymbol{r} \boldsymbol{\cdot} \boldsymbol{F}_1).
\label{eq:correctionFlowColdParticle}
\end{equation}
This flow field decays with the second power of the distance from particle 1 and is depicted in Fig. \ref{fig:correctionFlowHotParticle} for a local increase in viscosity. 
Along the direction of the force applied, the correction flow is parallel to the force, thus enhancing the constant viscosity Stokeslet flow. This is reasonable since with increased temperature the viscosity around the particle is decreased and the flow field becomes stronger. However, on the axis perpendicular to the force, the correction flow vanishes, concentrating the correction flow along the direction of the force applied. 
As a consistency check, we find that the divergence of the flow field \eqref{eq:correctionFlowColdParticle} vanishes and thus satisfies the incompressibility condition. 
In contrast to a linear viscosity gradient, this correction flow field decays stronger than the constant-viscosity Stokeslet by one order.

\section{Conclusion}
\label{ch:conclusion}
In this work, we have calculated the first order correction to the mobility matrix of spherical particles in a large linear viscosity gradient and in viscosity perturbations due to the particle temperature in the limit of low Peclet number. 
For a linear viscosity gradient, we have shown that the leading order correction to the interaction is independent of the distance between the particles, a result that is valid as long as the particle separation is small compared to the size of the viscosity gradient. The result found is in agreement with the generalised Oseen tensor derived in \citet{Laumann2019}. 
The increase in the exponent associated to the radial decay ultimately results from the viscosity being a linear field and therefore lifting the exponent by one in the corresponding integrals. 
Similarly, the flow field produced by a particle subject to a torque, which typically scales with the second inverse power of the distance to the particle, experiences a correction due to the viscosity gradient which decays slower by one order. The corrections for the mobility matrix presented here are valid not only for two-particle systems but extend, at the level of approximation chosen, to many particles since three-body and higher-order effects decay quicker than $s^{-1}$. Therefore, the results presented here are also applicable to dilute suspensions of many interacting particles.

In addition, we have shown that if either the particles are placed away from the centre of symmetry of the gradient or if the gradient lacks such symmetry, translational self-mobilities and interactions have to be corrected with respect to the symmetric case by an additional term. This feature is important since any physical viscosity gradient must be of finite size to avoid the pathological case of negative viscosity. 
This correction has to the best of our knowledge not been reported yet.  
It alters, in particular, the far field flow produced by a particle subject to an external force (Fig. \ref{fig:flowFieldAsymmetricallyPlacedParticle}) and becomes relevant already when the particles are placed away from the centre of symmetry by only a fraction of the gradient size. 
For asymmetric particle placement in an interface-like viscosity gradient, the resulting flow field is typically stronger towards the centre of symmetry and is weaker towards the edge of the viscosity gradient.

For particles at different temperatures than the surrounding fluid, we calculated the resulting temperature and viscosity fields in the fluid using a reflection method. We assumed that the Peclet number is small such that at length scales of the particle separation conductive effects dominate over advective effects. In this case, the self-mobilities of both particles are increased for hot and decreased for cold particles already at the zeroth order in $s$. At this order, our results are in exact agreement with the findings presented in \citet{Oppenheimer2016}. In addition, however, we found at the next higher order in the particle separation that the self-mobilities become higher or lower by the presence of a nearby particle with increased or decreased temperature relative to the surrounding fluid, respectively. Essentially, the second particle heats up or cools down the environment of the first one and alters its mobility. Finally, while for a single particle with a homogeneous surface temperature no coupling between translation and rotation is observed \citep{Oppenheimer2016}, such a term arises due to the presence of a second hot or cold particle and scales with the second inverse power of the particle separation 

The correction to the hydrodynamic interaction between two particles scales to leading with the second inverse power of the particle separation and is proportional to the sum of the temperature differences of both particles with respect to the fluid. Thus, whether the particle producing the flow or the particle sensing the flow has an increased or decreased temperature has no effect on the interaction between the two, which can be understood from the symmetry of the mobility matrix. 
In comparison to the constant-viscosity Stokeslet, this correction term decays quicker by one order in the particle separation, which coincides with the fact that the viscosity perturbation scales to leading order as $s^{-1}$. 

We believe that the results presented here may shed new light on many particulate systems featuring non-constant viscosity. A notable example is the behaviour of microswimmers in such an environment reigned by viscosity gradients. A main finding in \citet{Liebchen2018} was that uniaxial swimmers, i.e. particle-based swimmers with all beads on a common axis, do not adjust their swimming direction due to a viscosity gradient when the $\epsilon^1$-correction to the particle interaction is neglected. This essentially is because all forces act along the common axis such that this axis must be preserved in time. However, our results show that in such an arrangement of the particles, the first order correction to the interaction can induce a flow with a component orthogonal to the common axis when the axis itself is not aligned with the gradient (Fig. \ref{fig:flowField} (c)). Hence, the correction terms to particle interaction found here might allow for a rotation due to the viscosity gradient also for uniaxial swimmers, a result which would coincide with the recently reported viscotaxis for uniaxial squirmers \citep{Datt2019a}. 
Also it has been pointed out recently that the viscotactic behaviour of a squirmer in a viscosity gradient caused by e.g. a temperature field also depends on the temperature boundary condition at the squirmer surface \citep{Shaik2021}. In this context, the results presented in this work for hot particles become relevant in the collective behaviour of such microswimmers.
We plan to address the impact of our results on viscotaxis of microswimmers in a future publication. 

The results and methods presented here might also shed new light on the collective behaviour of Janus particles driven by laser light \citep{Buttinoni2012, Bauerle2018}, where the incident radiation leads to an increase in particle temperature and thus also to viscosity gradients around the particles. 
Besides the context of microswimmers, the results presented could also prove useful in sedimentation and centrifugation of nanoparticles, which often feature density and viscosity gradients \citep{Qiu2011, Pluisch2019}, or in diffusion of particles or polymers through a pore featuring a viscosity gradient \citep{DeHaan2013}. This holds in particular for systems with higher particle concentration, where interaction effects become relevant. Furthermore, viscosity gradients are abundant in biological environments with low Reynolds number \citep{Wilking2011, Swidsinski2007}, and the results presented here may lead to a better understanding of the interaction of active components therein.


\backsection[Acknowledgements]{The authors thank M. Hubert for valuable discussions.}

\backsection[Funding]{The authors acknowledge support by the German Research Foundation (DFG) within the Collaborative Research Center 1411 "Design of Particulate Products".}

\backsection[Declaration of interests]{The authors report no conflict of interest.}

\backsection[Data availability statement]{Data sharing is not applicable to this article as no new data were created or analysed in this study.}

\backsection[Author ORCID]{S. Ziegler https://orcid.org/0000-0002-1806-6230; A.-S. Smith, https://orcid.org/0000-0002-0835-0086}

\appendix

\section{Derivation of the Lorentz Reciprocal Theorem} \label{sec:app_1}
Finding the mobility and resistance matrix of an ensemble of particles in the general case of non-constant viscosity corresponds to solving the boundary value problem given by the Stokes equations (Eq. \eqref{eq:StokesEquation}) for the fluid velocity with the boundary conditions \citep{Oppenheimer2016},
\begin{equation}
\boldsymbol{u} (\boldsymbol{r}) = \boldsymbol{U}_b + \boldsymbol{\Omega}_b \times (\boldsymbol{r} - \boldsymbol{r}_b) \ \forall \boldsymbol{r} \in S_b, \ \ \boldsymbol{u}(\boldsymbol{r}) \to 0, p(\boldsymbol{r}) \to 0 \ \ \mathrm{for} \ \boldsymbol{r} \to \infty,
\label{eq:boundaryConditionsParticles}
\end{equation}
at the surfaces of the particles $S_b$ and at infinity, respectively. 
With prescribed (angular) particle velocities $\boldsymbol{U}_b$ and $\boldsymbol{\Omega}_b$, the hydrodynamic forces and torques acting on the particles are obtained by \citep{Oppenheimer2016} \begin{equation}
\boldsymbol{F}^\mathrm{hydro}_b = \int_{S_b} \boldsymbol{n} \boldsymbol{\cdot} \mathsfbilow{\sigma} dS, \ \boldsymbol{T}^\mathrm{hydro}_b = \int_{S_b} (\boldsymbol{r} - \boldsymbol{r}_b) \times (\boldsymbol{n} \boldsymbol{\cdot} \mathsfbilow{\sigma}) dS,
\label{eq:hydrodynamicForces}
\end{equation}
and yield the resistance matrix by means of Eq. \eqref{eq:defResistanceMatrix}. Here, $dS$ denotes the infinitesimal area element. With this expressions at hand, we exploit the Lorentz reciprocal theorem to derive the first order correction in $\epsilon$ to the mobility matrix.

We calculate following \citet{Oppenheimer2016},
\begin{eqnarray}
0 = && (\bnabla \boldsymbol{\cdot} \mathsfbilow{\sigma}) \boldsymbol{\cdot} \boldsymbol{u}^{(0)}  = \bnabla \boldsymbol{\cdot} (\mathsfbilow{\sigma} \boldsymbol{\cdot} \boldsymbol{u}^{(0)}) - \mathsfbilow{\sigma} : \bnabla \boldsymbol{u}^{(0)} = \nonumber \\ && \bnabla \boldsymbol{\cdot} (\mathsfbilow{\sigma} \boldsymbol{\cdot} \boldsymbol{u}^{(0)}) - \left( -p \mathsfbi{I} + 2 \eta(\boldsymbol{r}) \mathsfbi{E} \right) : \bnabla \boldsymbol{u}^{(0)} = \bnabla \boldsymbol{\cdot} (\mathsfbilow{\sigma} \boldsymbol{\cdot} \boldsymbol{u}^{(0)}) - 2 \eta(\boldsymbol{r}) \, \mathsfbi{E} : \mathsfbi{E}^{(0)},
\label{eq:LRT_Part1}
\end{eqnarray}
where in the last step we have used that $\mathsfbi{I} : \bnabla \boldsymbol{u}^{(0)} = 0$ due to incompressibility and ${\mathsfbi{E} : \bnabla \boldsymbol{u}^{(0)} = \mathsfbi{E} : \mathsfbi{E}^{(0)}}$ because $\mathsfbi{E}$ is a symmetric tensor. By '$:$' we denote the contraction of the last two indices of the first tensor with the first two indices of the second tensor, i.e. $\mathsfbi{E} : \mathsfbi{E}^{(0)} :=\mathsfbi{E}_{kl} \mathsfbi{E}^{(0)}_{kl}$. 

With the same argumentation we find that
\begin{equation}
0 = (\bnabla \boldsymbol{\cdot} \mathsfbilow{\sigma}^{(0)}) \boldsymbol{\cdot} \boldsymbol{u} = \bnabla \boldsymbol{\cdot} (\mathsfbilow{\sigma}^{(0)} \boldsymbol{\cdot} \boldsymbol{u}) - 2 \eta^{(0)} \, \mathsfbi{E} : \mathsfbi{E}^{(0)}. 
\label{eq:LRT_Part2}
\end{equation}
By subtraction of Eq. \eqref{eq:LRT_Part1} and Eq. \eqref{eq:LRT_Part2}, we obtain
\begin{equation}
\bnabla \boldsymbol{\cdot} (\mathsfbilow{\sigma} \boldsymbol{\cdot} \boldsymbol{u}^{(0)}) - \bnabla \boldsymbol{\cdot} (\mathsfbilow{\sigma}^{(0)} \boldsymbol{\cdot} \boldsymbol{u}) = 2 \left[ \eta(\boldsymbol{r}) - \eta^{(0)} \right] \mathsfbi{E} : \mathsfbi{E}^{(0)}. 
\label{eq:LRT}
\end{equation}
Under the assumption $\eta(\boldsymbol{r}) = \eta^{(0)}$ the right-hand side vanishes and we arrive at the Lorentz reciprocal theorem in its usual form.

Instead, we proceed by integrating Eq. \eqref{eq:LRT} over the whole fluid domain $V$. The left-hand side can then be rewritten using the divergence theorem as an integral over the bounding surface of the fluid. Since we assume fluid velocity and pressure to decay to zero at infinitely large distances from the particles, the integral over the bounding surface at infinity can be neglected \citep{Oppenheimer2016} and we only need to integrate over the particle surfaces: 
\begin{equation}
- \int_{S_1 \cup S_2} \boldsymbol{n} \boldsymbol{\cdot} \mathsfbilow{\sigma} \boldsymbol{\cdot} \boldsymbol{u}^{(0)} dS + \int_{S_1 \cup S_2} \boldsymbol{n} \boldsymbol{\cdot} \mathsfbilow{\sigma}^{(0)} \boldsymbol{\cdot} \boldsymbol{u} \, dS = \int_V 2 \left[ \eta(\boldsymbol{r}) - \eta^{(0)} \right] \mathsfbi{E} : \mathsfbi{E}^{(0)} dV. 
\end{equation}
Using Eq. \eqref{eq:externalForces} and Eqs. \eqref{eq:boundaryConditionsParticles} and \eqref{eq:hydrodynamicForces} we obtain Eq. \eqref{eq:intermediateStep}.

\section{Details on the contraction of the normalised strain rates} \label{sec:app_2}
The strain rates $\mathsfbi{E}$ and $\mathsfbi{E}^{(0)}$ are second rank tensors and hence the normalised strain rates $\mathsfbi{B}$ and $\mathsfbi{B}^{(0)}$ are third rank tensors with, using explicit index notation,
\begin{equation}
\mathsfbi{E}_{kl} = \sum_{P, b, i} \left( \mathsfbi{B}^P_b \right)_{kli} \left( \boldsymbol{\mathcal{F}}^P_b \right)_i,
\end{equation}
and similarly for the strain rate in the constant viscosity fluid $\mathsfbi{E}^{(0)}$. 
Eq. \eqref{eq:generalizedLRT} then explicitly reads
\begin{eqnarray}
\sum_{P, Q, b, c} &&\left[ \left( \boldsymbol{\mathcal{F}}^{(0) Q}_c \right)_j \left( \mathsfbilow{\mu}^{QP}_{cb} \right)_{ji} \left( \boldsymbol{\mathcal{F}}^P_b \right)_i - \left( \boldsymbol{\mathcal{F}}^{P}_b \right)_i \left( \mathsfbilow{\mu}^{(0) PQ}_{bc} \right)_{ij} \left( \boldsymbol{\mathcal{F}}^{(0) Q}_c \right)_j \right]  \nonumber \\&&= - 2 \int_V \left[ \eta (\boldsymbol{r}) - \eta^{(0)} \right] \sum_{P, Q, b, c} \left( \mathsfbi{B}^P_b \right)_{kli} \left( \boldsymbol{\mathcal{F}}^P_b \right)_i \left( \mathsfbi{B}^{(0) Q}_c \right)_{klj} \left( \boldsymbol{\mathcal{F}}^{(0) Q}_c \right)_j dV,
\end{eqnarray}
where the summation indices are already renamed suitably and we sum over repeated spatial indices $i, j, k, l$. Using the linearity on both sides in $\boldsymbol{\mathcal{F}}$ and $\boldsymbol{\mathcal{F}}^{(0)}$, we obtain
\begin{equation}
\left( \mathsfbilow{\mu}^{QP}_{cb} \right)_{ji} = \left( \mathsfbilow{\mu}^{PQ}_{bc} \right)_{ij} = \left( \mathsfbilow{\mu}^{(0) PQ}_{bc} \right)_{ij} - 2 \int_V \left[ \eta (\boldsymbol{r}) - \eta^{(0)} \right] \left( \mathsfbi{B}^P_b \right)_{kli} \left( \mathsfbi{B}^{(0) Q}_c \right)_{klj} dV,
\end{equation}
where we also made use of the symmetry of the mobility matrix in the first step \citep{Oppenheimer2016}. This is equivalent to Eq. \eqref{eq:correctionMobilityMatrix}. The symmetry of the mobility matrix is a direct result from the Lorentz reciprocal theorem and is independent of the viscosity field \citep{Jeffrey1984}. 

\section{Expansion of the strain rates using the reflection method} \label{sec:app_3}
For an introduction to the application of reflection methods in low Reynolds number hydrodynamics, we refer to \citet[Chapter 8]{KimKarilla1991}. 
The extent to which the reflection scheme has to be developed, i.e. the number of reflections necessary, depends on the desired accuracy of the result for the mobility matrix in terms of the inverse particle distance. We will develop the reflection scheme up to second order, i.e. such that the flow fields are correctly described up to order $\textit{O}(s^{-2})$. From the subsequent application to first a large linear viscosity gradient and second viscosity perturbations due to hot particles, it will be apparent that this level of approximation suffices to calculate the correction to the mobility matrix up to the orders $\textit{O}(s^{-1})$ and $\textit{O}(s^{-2})$, respectively.  

We first consider the expansion of $\mathsfbi{B}^{(0) t}_b$, corresponding to particle $b$ being subject to a force while the other particle $c$ is force- and torque-free.  
At the zeroth order of the reflection scheme, particle $b$ induces a fluid flow given by $\mathsfbi{T}^\mathrm{trans} (\boldsymbol{r} - \boldsymbol{r}_b) \boldsymbol{\cdot} \boldsymbol{F}_b /(6 \pi \eta^{(0)} a)$.
The normalised strain rate corresponding to this flow field is given explicitly by
\begin{eqnarray}
(\mathsfbi{P}^{(0) t})_{kli}(\boldsymbol{r})&& :=  \frac{1}{6 \pi \eta^{(0)} a} \frac{1}{2} [\del_k  (\mathsfbi{T}^\mathrm{trans})_{li} (\boldsymbol{r})   + \del_l (\mathsfbi{T}^\mathrm{trans})_{ki} (\boldsymbol{r})]   \\ = &&\frac{1}{6 \pi \eta^{(0)} a}  \left[ \frac{3 a r_i}{4 r^3} \right.  \left. \left( \delta_{kl} - 3 \frac{r_k r_l}{r^2} \right) + \frac{3 a^3}{4 r^5} \left[ r_i \left( -\delta_{kl} + 5 \frac{r_k r_l}{r^2} \right)  - r_k \delta_{li} - r_l \delta_{ki} \right] \right].\nonumber
\end{eqnarray}

At the position of the other particle $c$, this flow field is conveniently decomposed in a Taylor expansion, where higher derivatives of the flow field scale with higher powers of $1/s$.
Since particle $c$ is force-free, the constant, zeroth order term in the Taylor expansion, which scales as $s^{-1}$, induces advection of particle $c$ but not a disturbance flow field.
Similarly, the anti-symmetric component of the gradient of the flow field induces rotation of the torque-free particle $c$ but a zero disturbance flow. The only disturbance flow field up to $\textit{O}(s^{-2})$ is due to the strain rate associated with the gradient of the flow field \citep[p. 196]{KimKarilla1991}. 

For particle $c$ immersed in a flow with local strain rate $\mathsfbi{E}^\mathrm{local}$, the disturbance flow caused by particle $c$ is given as \citep[p. 278]{Dhont1996}
\begin{equation}
\boldsymbol{u}^\mathrm{dist}(\boldsymbol{r}) = - \frac{5}{2} \left[\left(\frac{a}{R} \right)^3 - \left(\frac{a}{R} \right)^5 \right] \left( \hat{R} \boldsymbol{\cdot} \mathsfbi{E}^\mathrm{local} \boldsymbol{\cdot} \hat{R} \right) \boldsymbol{R} - \left( \frac{a}{R} \right)^5 \mathsfbi{E}^\mathrm{local} \boldsymbol{\cdot} \boldsymbol{R} =: \mathsfbi{\Gamma}(\boldsymbol{R}) : \mathsfbi{E}^\mathrm{local},
\label{eq:definitionGamma}
\end{equation}
with $\boldsymbol{R} := \boldsymbol{r} - \boldsymbol{r}_c$, $R = |\boldsymbol{R}|$, $\hat{R}$ the unit vector along $\boldsymbol{R}$, and $\mathsfbi{\Gamma}$ a rank 3 tensor. 
The strain rate corresponding to this disturbance flow field, $\mathsfbi{E}^\mathrm{dist} (\boldsymbol{r})$, can then be expressed with respect to the original local strain rate $\mathsfbi{E}^\mathrm{local}$ as 
\begin{equation}
    \mathsfbi{E}^\mathrm{dist}(\boldsymbol{r}) = \mathsfbi{P}^{(0) s} (\boldsymbol{r} - \boldsymbol{r}_c) : \mathsfbi{E}^\mathrm{local},
\end{equation}
with 
\begin{eqnarray}
(\mathsfbi{P}^{(0) s})_{mijk}  (\boldsymbol{R}) && =  \frac{1}{2}  \left[ \del_m \mathsfbi{\Gamma}_{ijk} (\boldsymbol{R}) + \del_i \mathsfbi{\Gamma}_{mjk} (\boldsymbol{R}) \right]  \nonumber \\ = \frac{5}{2} R_i R_j && R_k R_m \left( 5 \frac{a^3}{R^7} - 7 \frac{a^5}{R^9} \right) - \frac{5}{2} \delta_{im} R_j R_k\left( \frac{a^3}{R^5} - \frac{a^5}{R^7} \right) -  \frac{5}{4} (R_i R_k \delta_{mj} \label{eq:Ps} \\ + && R_m R_k \delta_{ij} +  R_i R_j \delta_{mk} +  R_m R_j \delta_{ik}) \left( \frac{a^3}{R^5} - 2 \frac{a^5}{R^7} \right) - \frac{a^5}{2 R^5} (\delta_{ij} \delta_{mk} + \delta_{mj} \delta_{ik}). \nonumber 
\end{eqnarray}
From this, the normalised strain rate in the two particle system can be expanded as
\begin{equation}
\mathsfbi{B}^{(0) t}_b \equiv \mathsfbi{B}^{(0) t}_b (\boldsymbol{r}) = \mathsfbi{P}^{(0) t} (\boldsymbol{r} - \boldsymbol{r}_b) + \mathsfbi{P}^{(0) s} (\boldsymbol{r} - \boldsymbol{r}_c) : \mathsfbi{P}^{(0) t} (\boldsymbol{r}_c - \boldsymbol{r}_b) + \textit{O}(s^{-3}),
\end{equation} 
with $b \neq c$, since $\mathsfbi{P}^{(0) t} (\boldsymbol{r}_c - \boldsymbol{r}_b)$ is indeed the local normalised strain rate at the position of particle $c$. 

Secondly, we follow the same procedure for $\mathsfbi{B}^{(0) r}_b$, i.e. the normalised strain rate corresponding to the fluid flow when particle $b$ is subject to a torque while particle $c$ is force- and torque-free. Particle $b$ induces a rotlet which corresponds to the normalised strain rate 
\begin{eqnarray}
(\mathsfbi{P}^{(0) r})_{kli} (\boldsymbol{r}) &&  = \frac{1}{8 \pi \eta^{(0)} a^3}\frac{1}{2}  [\del_k  (\mathsfbi{T}^\mathrm{rot})_{li} (\boldsymbol{r}) +  \del_l (\mathsfbi{T}^\mathrm{rot})_{ki} (\boldsymbol{r})]  \nonumber \\   && =-\frac{1}{8 \pi \eta^{(0)} a^3} \frac{3 a^3}{2 r^5}  (r_k r_m  \epsilon_{mli} + r_l r_m \epsilon_{mki}). 
\end{eqnarray}
However, the rotlet decays already with the second inverse power of the distance from the particle and again the constant term in a Taylor expansion of this flow at the position of particle $c$ does not induce a disturbance flow. We conclude that
\begin{equation}
\mathsfbi{B}^{(0) r}_b \equiv \mathsfbi{B}^{(0) r}_b (\boldsymbol{r}) = \mathsfbi{P}^{(0) r} (\boldsymbol{r} - \boldsymbol{r}_b) + \textit{O}(s^{-3}). 
\end{equation}

\section{Calculation of the Q-terms for the linear viscosity gradient} \label{sec:app_4}
Generalising from the interface-like viscosity gradient, the following calculation is not restricted to this particular gradient, but applies to all viscosity perturbations of the form 
\begin{equation}
\eta^{(1)} (\boldsymbol{r}) := \eta^{\prime (1)} \left( \frac{\boldsymbol{r}}{d} \right) =: \eta^{\prime (1)} (\boldsymbol{r}'),
\label{eq:scalingPropertyGradient}
\end{equation}
with the assumptions that the viscosity perturbation is linear around the origin ($\eta^{\prime (1)} (\boldsymbol{r}') = \boldsymbol{H}^{\prime (1)} \boldsymbol{\cdot} \boldsymbol{r}'$ for $|\boldsymbol{r}'| < C_1$, with $C_1$ a constant of $\textit{O}(1)$), it is bounded by $C_2  \eta_0$, i.e. $|\eta^{\prime (1)} (\boldsymbol{r})| \leq C_2  \eta_0$ for all $\boldsymbol{r}'$, with $C_2$ a constant of $\textit{O}(1)$, and $\eta^{\prime (1)} (\boldsymbol{r}')$ is continuous. 
For the interface-like viscosity gradient with reference viscosity $\eta^{(0)} = \eta_0$, we find
\begin{equation}
\eta^{\prime (1)} (\boldsymbol{r}') := 
\begin{cases}
- \eta_0, &y' < -1 \\
(0, \eta_0, 0) \boldsymbol{\cdot} \boldsymbol{r}', &-1 \leq y' \leq 1\\
\eta_0, & y' > 1.
\end{cases}
\end{equation}
The constants $C_1$ and $C_2$ are introduced to allow also asymmetrical placement of the interacting particles in e.g. the interface-like gradient of total diameter $d$. In this case, the distance to the boundary of the linear gradient is smaller than $d$, but still of $\textit{O}(d)$ by assumption (Section \ref{ch:asymmetricLinearGradient}), and the following calculation is valid. 

\subsection{Extension of the viscosity gradient}
\label{sec:app_4.1}
The calculation of the $\mathsfbi{Q}^{(1)}$ terms (Eq. \eqref{eq:defQ}) involves integration over the whole fluid domain $V$, which is in our model $\mathbb{R}^3$ with the volumes of both spheres, $V_1$ and $V_2$, excluded,
\begin{equation}
\mathsfbi{Q}^{(1) PQ}_{bc} := - 2 \int_V \eta^{(1)}(\boldsymbol{r}) \, \mathsfbi{P}^{(0) P} (\boldsymbol{r} - \boldsymbol{r}_b) \doublecontract \mathsfbi{P}^{(0) Q} (\boldsymbol{r} - \boldsymbol{r}_c) \, dV.
\label{eq:defQTerm}
\end{equation}
We will in this subsection show that, under the assumptions made for the viscosity perturbation, the $\mathsfbi{Q}^{(1)}$ terms up to $\textit{O}(\epsilon^1 \kappa^1)$ can be calculated by 
\begin{equation}
\tilde{\mathsfbi{Q}}^{(1) PQ}_{bc} := - 2 \int_V \tilde{\eta}^{(1)} (\boldsymbol{r}) \, \mathsfbi{P}^{(0) P} (\boldsymbol{r} - \boldsymbol{r}_b) \doublecontract \mathsfbi{P}^{(0) Q} (\boldsymbol{r} - \boldsymbol{r}_c) \, dV,
\label{eq:QTermExtended}
\end{equation} 
with the extended viscosity perturbation $\tilde{\eta}^{(1)} (\boldsymbol{r}) := \boldsymbol{H} \boldsymbol{\cdot} \boldsymbol{r}$ for all $\boldsymbol{r} \in V$. This holds as long as the viscosity perturbation is anti-symmetric and both particles have distances from the centre of symmetry that are smaller than $d$ by orders of magnitude. If one of those two assumptions does not hold, one has to include a correction term to Eq. \eqref{eq:QTermExtended} as shown below. 

From the explicit expressions Eqs. \eqref{eq:Pt}, \eqref{eq:Pr} and Eq. \eqref{eq:Ps}, we know that $\mathsfbi{P}^{(0)t} (\boldsymbol{r})$ decays to leading order as $r^{-2}$, whereas $\mathsfbi{P}^{(0)r} (\boldsymbol{r})$ and $\mathsfbi{P}^{(0)s} (\boldsymbol{r})$ decay to leading order as $r^{-3}$. 
We expand all normalised strain rates in Eqs. \eqref{eq:defQTerm} and \eqref{eq:QTermExtended} in a Taylor series around a position $\boldsymbol{r}_0$ which satisfies $|\boldsymbol{r}_b- \boldsymbol{r}_0|, |\boldsymbol{r}_c - \boldsymbol{r}_0| \ll d$, i.e. is close to both particles compared to $d$. We obtain up to the first order in the Taylor expansion 
\begin{eqnarray}
\mathsfbi{P}^{(0) P} (\boldsymbol{r} - \boldsymbol{r}_b) = \mathsfbi{P}^{(0) P} (\boldsymbol{r} - \boldsymbol{r}_0) - \bnabla \mathsfbi{P}^{(0) P}|_{\boldsymbol{r}_0} \boldsymbol{\cdot} (\boldsymbol{r}_b - \boldsymbol{r}_0) + \textit{O}(\bnabla \bnabla \mathsfbi{P}^{(0) P}) \nonumber \\
= \mathsfbi{P}^{(0) P} (\boldsymbol{r} - \boldsymbol{r}_0) + 
\begin{cases}
\textit{O}(|\boldsymbol{r}- \boldsymbol{r}_0|^{-3}), \ \ P = t \\
\textit{O}(|\boldsymbol{r}- \boldsymbol{r}_0|^{-4}), \ \ P = r, s \\
\end{cases}. 
\end{eqnarray}
Thus, the contraction $\mathsfbi{P}^{(0) P} (\boldsymbol{r} - \boldsymbol{r}_b) \doublecontract \mathsfbi{P}^{(0) Q} (\boldsymbol{r} - \boldsymbol{r}_c)$ will scales to leading order as $|\boldsymbol{r}- \boldsymbol{r}_0|^{-4}$ only when $P = Q = t$, and to leading order as $|\boldsymbol{r}- \boldsymbol{r}_0|^{-5}$ or lower for all other possible cases of $P, Q$. Moreover for $P = Q = t$, the $|\boldsymbol{r}- \boldsymbol{r}_0|^{-4}$ term is given explicitly as 
\begin{eqnarray}
&&\mathsfbi{P}^{(0) t} (\boldsymbol{r}- \boldsymbol{r}_b) \doublecontract \mathsfbi{P}^{(0) t} (\boldsymbol{r} - \boldsymbol{r}_c)  = \mathsfbi{P}^{(0) t} (\boldsymbol{r} - \boldsymbol{r}_0) \doublecontract \mathsfbi{P}^{(0) t} (\boldsymbol{r} - \boldsymbol{r}_0) + \textit{O}(|\boldsymbol{r} - \boldsymbol{r}_0|^{-5})  \nonumber \\ && = \frac{1}{(6 \pi a \, \eta^{(0)})^2} \frac{27 a^2}{8 |\boldsymbol{r} - \boldsymbol{r}_0|^4} \frac{(\boldsymbol{r} - \boldsymbol{r}_0)  \otimes (\boldsymbol{r} - \boldsymbol{r}_0)}{|\boldsymbol{r} - \boldsymbol{r}_0|^2} + \textit{O}(|\boldsymbol{r} - \boldsymbol{r}_0|^{-5}).
\label{eq:integrandCorrectionAsymmetry}
\end{eqnarray}

The error made by expanding the viscosity gradient to infinity is quantified as
\begin{equation}
\mathsfbi{Q}^{(1) PQ}_{bc} - \tilde{\mathsfbi{Q}}^{(1) PQ}_{bc} = - 2 \int_V \left[ \eta^{(1)} (\boldsymbol{r}) - \tilde{\eta}^{(1)} (\boldsymbol{r}) \right] \, \mathsfbi{P}^{(0) P} (\boldsymbol{r} - \boldsymbol{r}_b) \doublecontract \mathsfbi{P}^{(0) Q} (\boldsymbol{r} - \boldsymbol{r}_c) \, dV.
\label{eq:integralCorrectionAsymmetry}
\end{equation} 
Here, $\eta^{(1)} (\boldsymbol{r}) - \tilde{\eta}^{(1)} (\boldsymbol{r})$ is by definition zero for $|\boldsymbol{r}| < C_1 d$. Since the distances of both particles from the boundary of the homogeneous linear gradient is $\textit{O}(d)$, all values of $\boldsymbol{r}$ contributing to the integral \eqref{eq:integralCorrectionAsymmetry} due to non-zero $\eta^{(1)} (\boldsymbol{r}) - \tilde{\eta}^{(1)} (\boldsymbol{r})$ satisfy $|\boldsymbol{r} - \boldsymbol{r}_0| = \textit{O}(d)$. 
We thus express the integral \eqref{eq:integralCorrectionAsymmetry} in terms of $\boldsymbol{r}' := \boldsymbol{r}/d$ and $\boldsymbol{r}_0^\prime := \boldsymbol{r}_0/d$ and find that 
\begin{equation}
    \left[ \eta^{(1)} (\boldsymbol{r}) - \tilde{\eta}^{(1)} (\boldsymbol{r}) \right] = \eta^{\prime (1)} (\boldsymbol{r}') - \boldsymbol{H}^{(1)} \boldsymbol{\cdot} \boldsymbol{r} = \eta^{\prime (1)} (\boldsymbol{r}') - \boldsymbol{H}^{\prime (1)} \boldsymbol{\cdot} \boldsymbol{r}'.
\end{equation}
Thus, in the transformed integral the factor $\eta^{\prime (1)} (\boldsymbol{r}') - \boldsymbol{H}^{\prime (1)} \boldsymbol{\cdot} \boldsymbol{r}'$ is independent of $d$, while the leading order term in  Eq. \eqref{eq:integrandCorrectionAsymmetry} factors out a $d^{-4}$ and the volume element yields a factor $d^3$. Apart from the overall factor $d^{-1} \sim \kappa$, the integral \eqref{eq:integralCorrectionAsymmetry} is independent of $d$ and can be calculated, depending on $\eta^{\prime (1)} (\boldsymbol{r}')$, either numerically or also analytically. 
All contributions from the higher order terms in Eq. \eqref{eq:integrandCorrectionAsymmetry} scale lower in $d$ and thus can be neglected since we are interested only in the correction linear in $\kappa$. 
However, assuming that the viscosity perturbation is anti-symmetric with respect to $\boldsymbol{r} \to -\boldsymbol{r}$ and that $|\boldsymbol{r}_b|, |\boldsymbol{r}_c| \ll d$, which means we can choose $\boldsymbol{r}_0 = 0$, we find that the integral \eqref{eq:integralCorrectionAsymmetry} vanishes at this order $\textit{O}(d^{-1})$ due to symmetry. 

We conclude that for anti-symmetric viscosity gradients and both particles close to the centre of symmetry, extending the viscosity gradient to infinity leaves indeed the $\kappa^1$ component of all $\mathsfbi{Q}^{(1)}$ terms untouched, which simplifies the subsequent calculation of the $\mathsfbi{Q}^{(1)}$ terms. If one of these conditions is not satisfied, we generally calculate the $\mathsfbi{Q}^{(1)}$ terms with the extended viscosity gradient, but will have to account for an additional contribution to all $\mathsfbi{Q}^{(1) tt}$ terms given by
\begin{equation}
\mathsfbi{Q}^{(1) tt}_\mathrm{asym} = - \frac{2}{(6 \pi a \, \eta^{(0)})^2} \int_V \left[ \eta^{(1)} (\boldsymbol{r}) - \tilde{\eta}^{(1)} (\boldsymbol{r}) \right] \frac{27 a^2}{8 |\boldsymbol{r} - \boldsymbol{r}_0|^6} (\boldsymbol{r} - \boldsymbol{r}_0) \otimes (\boldsymbol{r} - \boldsymbol{r}_0) \, dV.
\end{equation}
Importantly, this correction applies to both the $\mathsfbi{Q}^{(1) tt}_{bc}$ terms with $b = c$ (self terms) and well as $b \neq c$ (cross terms) in similar fashion. 

\subsection{Self Q terms}
\label{sec:app_4.2}
\begin{figure}
\begin{center}
\includegraphics[scale=0.5]{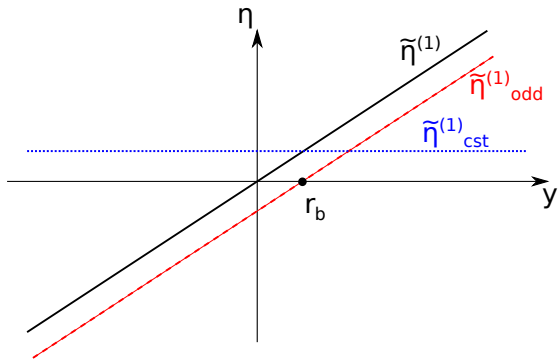}
\caption{Sketch of the decomposition of the viscosity gradient.}
\label{fig:sketchDecompositionEta}
\end{center}
\end{figure}

In order to calculate self terms of the form of Eq. \eqref{eq:QTermExtended} with $b = c$, we decompose the extended viscosity gradient as
\begin{equation}
\tilde{\eta}^{(1)} (\boldsymbol{r}) = \tilde{\eta}^{(1)} (\boldsymbol{r}_b) + \tilde{\eta}^{(1)}(\boldsymbol{r} - \boldsymbol{r}_b) =: \tilde{\eta}^{(1)}_\mathrm{cst} + \tilde{\eta}^{(1)}_\mathrm{odd} (\boldsymbol{r} - \boldsymbol{r}_b). 
\end{equation}
This decomposition is sketched in Fig. \ref{fig:sketchDecompositionEta}. 
Both $\tilde{\eta}^{(1)}_\mathrm{cst}$ and $\tilde{\eta}^{(1)}_\mathrm{odd} (\boldsymbol{r} - \boldsymbol{r}_b)$ as well as the contracted normalised strain rates can be expressed in spherical coordinates with the origin in $\boldsymbol{r}_b$. 
We find \citep{Oppenheimer2016}
\begin{equation}
\left( \mathsfbi{P}^{(0) t} (\boldsymbol{r} - \boldsymbol{r}_b) \doublecontract \mathsfbi{P}^{(0) t} (\boldsymbol{r} - \boldsymbol{r}_b) \right)_{ij} = \frac{1}{(6 \pi \eta^{(0)} a)^2} \frac{9 a^6}{8 r^8}  \left(\left[3 \left( \frac{r}{a}\right)^4 - 6 \left(\frac{r}{a}\right)^2 + 2\right] \frac{r_i r_j}{r^2} + \delta_{ij} \right), 
\label{eq:Sttbb}
\end{equation}
\begin{eqnarray}
\left( \mathsfbi{P}^{(0) t} (\boldsymbol{r} - \boldsymbol{r}_b) \doublecontract \mathsfbi{P}^{(0) r} (\boldsymbol{r} - \boldsymbol{r}_b) \right)_{ij} &&= \left( \mathsfbi{P}^{(0) r} (\boldsymbol{r} - \boldsymbol{r}_b) \doublecontract \mathsfbi{P}^{(0) t} (\boldsymbol{r} - \boldsymbol{r}_b) \right)_{ji} \nonumber \\ &&= \frac{1}{(6 \pi \eta^{(0)} a)^2} \frac{9 a^6}{4 r^8} \epsilon_{ijk} r_k, 
\label{eq:Strbb}
\end{eqnarray}
\begin{equation}
\left( \mathsfbi{P}^{(0) r} (\boldsymbol{r} - \boldsymbol{r}_b) \doublecontract \mathsfbi{P}^{(0) r} (\boldsymbol{r} - \boldsymbol{r}_b) \right)_{ij} = \frac{1}{(6 \pi \eta^{(0)} a)^2} \frac{9 a^6}{2 r^6} \left( \delta_{ij} - \frac{r_i r_j}{r^2} \right).
\label{eq:Srrbb}
\end{equation}

In order to perform the integration \eqref{eq:QTermExtended}, it is convenient to include the volume of the other particle into the integration domain, such that the integral runs over $r \in [a, \infty], \theta \in [0, \pi]$ and $\varphi \in [0, 2 \pi]$. 
Since the contracted normalised strain rates decay at least as $r^{-4}$ or faster and the viscosity gradient scales to leading order as $r^1$, with $r$ the distance from the particle, the error made by including the volume of the other particle scales at most as $s^{-3}$ and can be neglected at our level of approximation. 
We obtain by straight-forward integrations
\begin{gather}
\mathsfbi{Q}^{(1) tt}_{bb} = - \frac{1}{6 \pi a (\eta^{(0)})^2} \, \eta^{(1)} (\boldsymbol{r}_b) \mathsfbi{I} + \textit{O}(s^{-3}), \nonumber \\
\mathsfbi{Q}^{(1) rr}_{bb} = - \frac{1}{8 \pi a^3 (\eta^{(0)})^2} \, \eta^{(1)} (\boldsymbol{r}_b) \mathsfbi{I} + \textit{O}(s^{-5}), \label{eq:QTermsLinGradSelfMob}\\
\mathsfbi{Q}^{(1) tr}_{bb} = \frac{1}{24 \pi a (\eta^{(0)})^2} (\bnabla \eta^{(1)} \times) + \textit{O}(s^{-4}), \nonumber 
\end{gather}
with $(\bnabla \eta^{(1)} \times)_{ij} := \del_k \eta^{(1)} \epsilon_{ikj}$. Note that since both particles are well within the domain where $\eta^{(1)}$ is homogeneously linear, we can use the original viscosity variation instead of the extended one in the above expressions. 

\subsection{Cross Q terms}
\label{sec:app_4.3}
To simplify the calculation of the cross terms $\mathsfbi{Q}^{(1) PQ}_{bc}$ with $b \neq c$ and $P, Q = t, r$ using Eq. \eqref{eq:QTermExtended}, we firstly show that, being precise up to order $s^{-1}$, we can include the volumes of both particles in the integration domain without affecting the result of the integral. 
To estimate the error made by including the volume of without restriction of generality particle $b$ in the integration domain, we use a Taylor expansion of the strain rate corresponding to the other particle $c \neq b$ around $\boldsymbol{r}_b$. 
If the translation/rotation mode associated to particle $c$ is $Q = r$ or $Q = s$, the corresponding strain rate at $\boldsymbol{r}_b$ will scale as $s^{-3}$ to leading order. Since the viscosity variation scales to leading order linear and thus at most as $s^1$, the corresponding integral over the domain of particle $b$ then scales at most as $s^{-2}$ and can be neglected. 

If $Q = t$, the translational strain rate corresponding to particle $c$ will scale to leading order as $s^{-2}$ at $\boldsymbol{r}_b$. All higher order terms in the expansion scale at least as $s^{-3}$ and thus can be neglected similarly to before. For the case $P = t$, the strain rate due to translation of particle $b$ is anti-symmetric under inversion about $\boldsymbol{r}_b$. Since it is contracted with the leading order term of the expansion of the product of viscosity variation times the normalised strain rate corresponding to particle $c$, which is a constant near $\boldsymbol{r}_b$, the integration of the $s^{-2}$ term over the domain of particle $b$ vanishes due to symmetry. 
Finally, if $Q = t$ and $P = r$, we calculate explicitly the contraction of $\mathsfbi{P}^{(0) r} (\boldsymbol{r} - \boldsymbol{r}_b)$ with the leading order term in an expansion of $\mathsfbi{P}^{(0) t} (\boldsymbol{r} - \boldsymbol{r}_c)$ around $\boldsymbol{r}_b$. Assuming without restriction of generality $b = 2$ and $c = 1$, we obtain for the expansion around $\boldsymbol{r}_2 = 0$
\begin{eqnarray}
\left(\mathsfbi{P}^{(0) r} (\boldsymbol{r} - \boldsymbol{r}_2) \doublecontract \mathsfbi{P}^{(0) t} (\boldsymbol{r} - \boldsymbol{r}_1)\right)_{ij} && = \left( \mathsfbi{P}^{(0) r} (\boldsymbol{r}) \doublecontract \mathsfbi{P}^{(0) t} (\boldsymbol{s}) \right)_{ij} + \textit{O}(s^{-3})  \nonumber \\ = \mathsfbi{P}^{(0) r}_{kli} (\boldsymbol{r}) \mathsfbi{P}^{(0) t}_{klj} (\boldsymbol{s}) + \textit{O}(s^{-3})  \nonumber \\
=- \frac{1}{48 \pi^2 (\eta^{(0)})^2}  \frac{9 s_j}{8 r^5 s^3} && \left(\delta_{kl} - 3 \frac{s_k s_l}{s^2}  \right) \left(r_k r_m \epsilon_{mli} + r_l r_m \epsilon_{mki} \right) + \textit{O}(s^{-3})  = \label{eq:expansionRTTerm} \\
&& \frac{1}{48 \pi^2 (\eta^{(0)})^2} \frac{9 s_j}{8 r^5 s^3}  3 \frac{s_k s_l}{s^2}  \left(r_k r_m \epsilon_{mli} + r_l r_m \epsilon_{mki} \right) + \textit{O}(s^{-3}), \nonumber
\end{eqnarray}
where in the last step we have used the anti-symmetry of $\epsilon_{ijk}$, the Levi-Civita tensor. 
Assuming without loss of generality that $\boldsymbol{s} = (0, 0, s)$, we find that the above contraction is anti-symmetric with respect to $z \to -z$. Namely, for any non-zero contribution of Eq. \eqref{eq:expansionRTTerm}, both $k$ and $l$ must correspond to the $z$-direction as otherwise the $s_k s_l$ vanishes. Due to the anti-symmetry of the $\epsilon$ tensor, $m$ cannot be associated to the $z$-direction, which proofs the anti-symmetry. The leading order term in an expansion of the viscosity perturbation is a constant and hence the integral of Eq. \eqref{eq:expansionRTTerm} times the viscosity perturbation over the volume of particle $b$ vanishes up to order $\textit{O}(s^{-1})$ due to symmetry. This proofs our claim. 

To calculate the cross terms, we secondly decompose the viscosity gradient with respect to the middle between both particles $\boldsymbol{r}_m = (\boldsymbol{r}_1 + \boldsymbol{r}_2)/2$ as
\begin{equation}
\tilde{\eta}^{(1)} (\boldsymbol{r}) = \tilde{\eta}^{(1)} (\boldsymbol{r}_m) + \tilde{\eta}^{(1)}(\boldsymbol{r} - \boldsymbol{r}_m) =: \tilde{\eta}^{(1)}_\mathrm{cst} + \tilde{\eta}^{(1)}_\mathrm{odd} (\boldsymbol{r} - \boldsymbol{r}_m). 
\end{equation}
Without restriction of generality, we assume that the particles are positioned at $\boldsymbol{r}_1 = (0, 0, -s/2)$ and $\boldsymbol{r}_2 = (0, 0, s/2)$. We employ bispherical coordinates, defined as \citep{Moon1988}
\begin{equation}
x = \frac{s}{2} \frac{\sin \sigma \cos \phi}{\cosh \tau - \cos \sigma}, \ y = \frac{s}{2} \frac{\sin \sigma \sin \phi}{\cosh \tau - \cos \sigma}, \ z = \frac{s}{2} \frac{\sinh \tau}{\cosh \tau - \cos \sigma},
\end{equation}
with integration boundaries given as $\tau \in (-\infty, \infty), \sigma \in [0, \pi]$ and $\phi \in [0, 2 \pi]$ corresponding to the complete $\mathbb{R}^3$. 
The integration boundaries are independent of $s$, so in order to obtain the integral up to $\textit{O}(s^{-1})$ it suffices to perform the corresponding Taylor expansion on the integrand and integrate only terms up to $\textit{O}(s^{-1})$. 
Since this expansion and the subsequent calculation are tedious, we employ Mathematica \citep{Mathematica2020} to carry them out. 
We finally obtain 
\begin{eqnarray}
\mathsfbi{Q}^{(1) tt}_{12} = \left( \mathsfbi{Q}^{(1) tt}_{21} \right)^T &&  \nonumber \\ = \frac{1}{6 \pi a (\eta^{(0)})^2} \, && \left( - \eta^{(1)} (\frac{\boldsymbol{r}_1 + \boldsymbol{r}_2}{2}) \, \mathsfbi{T} + \frac{3 a}{8 s} \left[ \bnabla \eta^{(1)} \otimes \boldsymbol{s} - \boldsymbol{s} \otimes \bnabla \eta^{(1)} \right] \right) + \textit{O}(s^{-2}), \nonumber \\
\mathsfbi{Q}^{(1) rr}_{12} = \mathsfbi{Q}^{(1) rr}_{21} = \textit{O}(&&s^{-2}) , \label{eq:QTermLinGradInteraction}  
\\
\mathsfbi{Q}^{(1) tr}_{12} = \left( \mathsfbi{Q}^{(1) rt}_{21} \right)^T = && \frac{1}{6 \pi a (\eta^{(0)})^2} \left( - \eta^{(1)} \left( \frac{\boldsymbol{r}_1 + \boldsymbol{r}_2}{2} \right) \mathsfbi{V} + \frac{3 a}{8 s^3} \boldsymbol{s} \otimes (\boldsymbol{s} \times \bnabla \eta^{(1)}) \right) + \textit{O}(s^{-3}), \nonumber 
\\
\mathsfbi{Q}^{(1) tr}_{21} = \left( \mathsfbi{Q}^{(1) rt}_{12} \right)^T = && \frac{1}{6 \pi a (\eta^{(0)})^2} \left( +\eta^{(1)} \left( \frac{\boldsymbol{r}_1 + \boldsymbol{r}_2}{2} \right) \mathsfbi{V} + \frac{3 a}{8 s^3} \boldsymbol{s} \otimes (\boldsymbol{s} \times \bnabla \eta^{(1)}) \right) + \textit{O}(s^{-3}).\nonumber
\end{eqnarray}
Here, $^T$ denotes matrix transposition. 

\subsection{Proof of Eq. \eqref{eq:simpleMuLinearGradient}}
\label{sec:app_4.4}
We first consider the case of $b = c$ in Eqs. \eqref{eq:muInTermsOfQ}. With $d \neq b$ and $e \neq c$, we find that $e = d \neq b = c$ and thus all $\mathsfbi{Q}^{(1)}$ terms involving an upper index $s$ are cross terms. 
However, cross terms with an upper index $s$ scale at most as $s^{-1}$: First, including the volumes of the particles in the integration domain implies an error of at most $\textit{O}(s^{-1})$ since each normalised strain rate decays at least as $s^{-2}$ and the viscosity gradient scales to leading order only as $s^1$, such that the integrand scales at most as $s^{-1}$ in the vicinity of each particle. The remaining integral with the particle volumes included also scales at most as $s^{-1}$ because the normalised strain rates, expressed in bispherical coordinates as described above, scale as $s^{-3}$ ($\mathsfbi{P}^{(0) s}$) and at most as $s^{-2}$ (the remaining one if it is associated with translation), the viscosity gradient scales linear in $s$ and the volume element of bispherical coordinates as $s^3$. 
Since this cross term is contracted to a normalised strain rate in Eqs. \eqref{eq:muInTermsOfQ}, which decays at least as $s^{-2}$, those contractions can be neglected. 

For the case $b \neq c$ we find that $e = b \neq c = d$ and all $\mathsfbi{Q}^{(1)}$ terms involving an upper index $s$ are self terms. Self terms of the form $\mathsfbi{Q}^{(1)ts}_{bb}$ or $\mathsfbi{Q}^{(1)st}_{bb}$ scale to leading order as $s^0$ since they only depend on $\tilde{\eta}^{(1)}_\mathrm{odd} (\boldsymbol{r} - \boldsymbol{r}_b)$ due to symmetries of the $\mathsfbi{Q}$ terms, which can be seen from explicit calculations. Their contraction to a normalised strain rate thus scales at most as $s^{-2}$ and can be neglected. Terms of the form $\mathsfbi{Q}^{(1)sr}_{bb}$ are anti-symmetric in all three indices, which can be verified by direct calculation, and thus vanish when two of those indices are contracted to a symmetric normalised strain rate $\mathsfbi{P}^{(0) t}$. 

\subsection{Correction of the Q terms for particles placed asymmetrically in the interface-like gradient}
\label{sec:app_4.5}
\begin{figure}
\begin{center}
\includegraphics[scale=0.5]{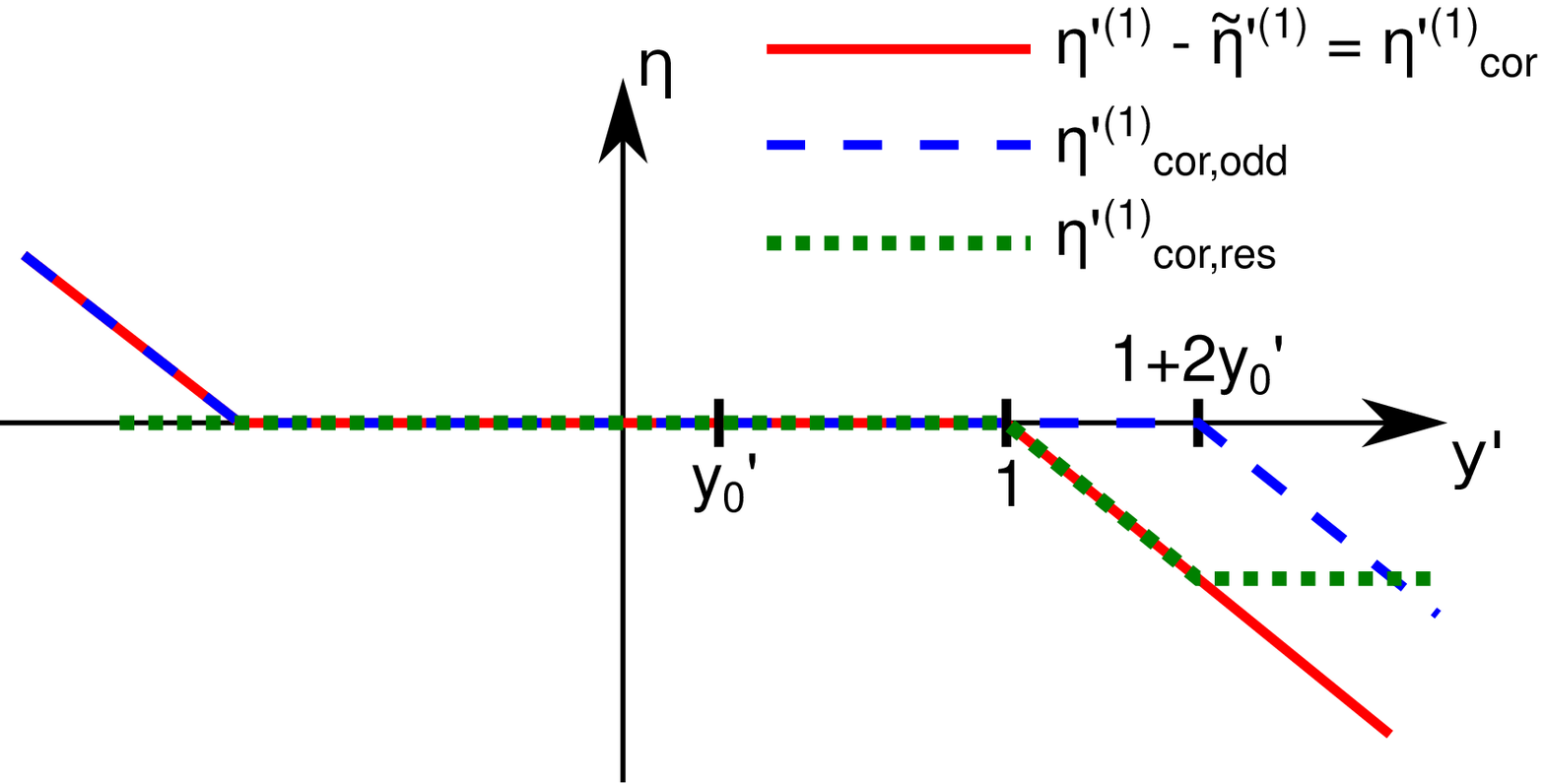}
\caption{Sketch of the decomposition of $\eta^{\prime (1)}_\mathrm{cor}$ for $y_0' > 0$.}
\label{fig:sketchDecompositionEtaCor}
\end{center}
\end{figure}

In order to calculate Eq. \eqref{eq:correctionAsymmetry}, we first assume that without restriction of generality $\boldsymbol{r}_0 = (0, y_0, 0)$ exploiting the symmetry of the interface-like gradient along the $x$- and $z$-direction. 
Introducing rescaled coordinates as in Section \ref{sec:app_4.1} and $y'_0 = y_0/d$, we find that
\begin{equation}
\mathsfbi{Q}^{(1) tt}_\mathrm{asym} = - \frac{2}{(6 \pi a \, \eta^{(0)})^2} \int_V \left[ \eta^{\prime (1)} (\boldsymbol{r}) - \boldsymbol{H}^{\prime (1)} \boldsymbol{\cdot} \boldsymbol{r}' \right] \frac{27 a^2}{8 d |\boldsymbol{r}' - \boldsymbol{r}'_0|^6} (\boldsymbol{r}' - \boldsymbol{r}'_0) \otimes (\boldsymbol{r}' - \boldsymbol{r}'_0) \, dV',
\label{eq:QTermCorrection}
\end{equation}
with $dV' = dx' dy' dz'$ the volume element in rescaled Cartesian coordinates. 
Since the contraction $\mathsfbi{P}^{(0) t} (\boldsymbol{r} - \boldsymbol{r}_0) \doublecontract \mathsfbi{P}^{(0) t} (\boldsymbol{r} - \boldsymbol{r}_0)$ is at the leading order $\textit{O}(|\boldsymbol{r} - \boldsymbol{r}_0|^{-4})$ symmetric under inversion with respect to $\boldsymbol{r}'_0$, it is convenient to decompose the term 
\begin{equation}
\eta^{\prime (1)} (\boldsymbol{r}') - \boldsymbol{H}^{\prime (1)} \boldsymbol{\cdot} \boldsymbol{r}' =: \eta'_\mathrm{cor} (\boldsymbol{r}')
\end{equation}
into a component $\eta^{\prime (1)}_\mathrm{cor, odd} (\boldsymbol{r}')$, which is anti-symmetric with respect to $\boldsymbol{r}'_0$, and into a residual component, $\eta^{\prime (1)}_\mathrm{cor, res} (\boldsymbol{r}') := \eta^{\prime (1)}_\mathrm{cor} (\boldsymbol{r}') - \eta^{\prime (1)}_\mathrm{cor, odd} (\boldsymbol{r}')$ (Fig. \ref{fig:sketchDecompositionEtaCor}). 

With 
\begin{equation}
\eta^{\prime (1)}_\mathrm{cor} (\boldsymbol{r}') = 
\begin{cases}
- H^{\prime (1)}_y (y' + 1), &y' < -1 \\
0, &-1 \leq y' \leq 1\\
- H^{\prime (1)}_y (y' - 1), & y' > 1,
\end{cases}
\end{equation}
a possible definition for the anti-symmetric component in the case of $y'_0 > 0$ is
\begin{equation}
\eta^{\prime (1)}_\mathrm{cor, odd} (\boldsymbol{r}') = 
\begin{cases}
- H^{\prime (1)}_y (y' + 1), &y' < -1 \\
0, &-1 \leq y' \leq 1 + 2 y'_0\\
- H^{\prime (1)}_y (y' - 1 - 2y'_0), & y' > 1 + 2 y'_0,
\end{cases}
\end{equation}
leaving the residual component as 
\begin{equation}
\eta'_\mathrm{cor, res} (\boldsymbol{r}') = 
\begin{cases}
0, &y' < 1 \\
-H^{\prime (1)}_y (y' -1), &1 \leq y' \leq 1 + 2 y'_0\\
- 2 H^{\prime (1)}_y y'_0, & y' > 1 + 2 y'_0.
\end{cases}
\end{equation}
Here we have assumed $\boldsymbol{H}^{\prime (1)} = (0, H^{\prime (1)}_y, 0)$.
Due to symmetry, the contribution of $\eta'_\mathrm{cor, odd}$ to the integral \eqref{eq:QTermCorrection} vanishes and it remains to calculate the contribution from $\eta'_\mathrm{cor, res}$. 
We obtain \citep{Mathematica2020}
\begin{equation}
-2 \int_V \eta^{\prime (1)}_\mathrm{cor, res} (\boldsymbol{r}') \frac{1}{|\boldsymbol{r}' - \boldsymbol{r}'_0|^6} (\boldsymbol{r}' - \boldsymbol{r}'_0) \otimes (\boldsymbol{r}' - \boldsymbol{r}'_0) \, dV' = \pi \arctanh(y'_0) \left( \mathsfbi{I} +  \hat{y} \otimes \hat{y} \right),
\end{equation}
and from this 
\begin{equation}
\mathsfbi{Q}^{(1) tt}_\mathrm{asym} = \frac{1}{6 \pi a (\eta^{(0)})^2} \frac{9 a |\bnabla \eta^{(1)}|}{16} \arctanh(y'_0) \left( \mathsfbi{I} +  \hat{y} \otimes \hat{y} \right). 
\label{eq:resultCorrectionAsymmetry}
\end{equation}
A similar calculation for the case $y'_0 < 0$ shows that Eq. \eqref{eq:resultCorrectionAsymmetry} is also valid in this case.

\section{Calculation of the Q terms for hot particles} \label{sec:app_5}
\subsection{Self Q terms} \label{sec:app_5.1}
We firstly show that, in the case of hot or cold particles, for a self $\mathsfbi{Q}^{(1)}$ term of the form 
\begin{equation}
\mathsfbi{Q}^{(1) PQ}_{bb} := - 2 \int_V \eta^{(1)}(\boldsymbol{r}) \, \mathsfbi{P}^{(0) P} (\boldsymbol{r} - \boldsymbol{r}_b) \doublecontract \mathsfbi{P}^{(0) Q} (\boldsymbol{r} - \boldsymbol{r}_b) \, dV,
\label{eq:HotPartSamePartTerm}
\end{equation}
we can include the volume of the other particle $c$ in the integration domain $V$ without affecting the result up to order $\textit{O}(s^{-2})$.
Assuming that the viscosity perturbation is of the form of Eq. \eqref{eq:viscosityPerturbationHotPart}, we find that the integrand in Eq. \eqref{eq:HotPartSamePartTerm} can be Taylor expanded around the position of particle $c$ and scales to leading order as $s^{-4}$, since both normalised strain rates decay at least with the second power in the inverse distance. The viscosity perturbation consists of a term centred around the first particle $b$ which scales to leading order as $s^{-1}$ around particle $c$, and a term centred around particle $c$ which has no $s$ dependence when evaluated locally. Thus, the error made by including the volume of particle $c$ in the integration domain scales as $s^{-4}$ or lower and can be neglected. 

Additionally, the contribution to the self $\mathsfbi{Q}^{(1)}_{bb}$ term from the dipolar viscosity field around particle $c$, i.e. proportional to $\boldsymbol{\eta}^{(1) d}_c$, can be neglected. 
This is because, firstly $\boldsymbol{C}^d_2$ scales as $s^{-2}$, and secondly because the corresponding integral without the prefactor also scales at most as the $s^{-2}$: The volumes of both particles can be included in the integration domain without affecting the result at order $\textit{O}(s^{-1})$ and from a scaling argument in $s$ using bispherical coordinates we conclude that this integral scales at most as $s^{-3}$. 

We then employ a spherical coordinate system around $\boldsymbol{r}_b$ to calculate the self $\mathsfbi{Q}^{(1)}$ terms. Since we need to account also for the monopolar viscosity perturbation around particle $c \neq b$ and the integration thus is not trivial, we use Mathematica \citep{Mathematica2020} for its computation. We obtain 
\begin{eqnarray}
&&\mathsfbi{Q}^{(1)tt}_{bb} = \frac{1}{6 \pi a (\eta^{(0)})^2} \left[ \left( -\frac{5 \eta^{(1) m}_b}{12} -\frac{a}{s} \eta^{(1) m}_c \left(1 - \frac{9a}{8s} \right) \right) \mathsfbi{I} - \frac{9 a^2}{8} \eta^{(1) m}_c \frac{\boldsymbol{s} \otimes \boldsymbol{s}}{s^4} \right] + \textit{O}(s^{-3}), \nonumber \\
&&\mathsfbi{Q}^{(1)rr}_{bb} = \frac{1}{6 \pi a (\eta^{(0)})^2} \left[-\frac{9}{16 a^2} \eta^{(1) m}_b - \frac{3}{4 as} \eta^{(1) m}_c \right] \mathsfbi{I} + \textit{O}(s^{-3}), \label{eq:QTermsHotPart} \\
&&\mathsfbi{Q}^{(1)tr}_{bb} = \frac{1}{6 \pi a (\eta^{(0)})^2} \left[ \frac{(\boldsymbol{\eta}^{(1) d}_b \times)}{8 a} + \frac{\eta^{(1) m}_c a}{4} \frac{(\boldsymbol{r}_c - \boldsymbol{r}_b)\times }{|\boldsymbol{r}_c - \boldsymbol{r}_b|^3}\right] + \textit{O}(s^{-3}), b \neq c. \nonumber
\end{eqnarray}

\subsection{Cross terms}\label{sec:app_5.2}
In this section we describe the calculation of the cross terms 
\begin{equation}
\mathsfbi{Q}^{(1) PQ}_{bc} := - 2 \int_V \eta^{(1)}(\boldsymbol{r}) \, \mathsfbi{P}^{(0) P} (\boldsymbol{r} - \boldsymbol{r}_b) \doublecontract \mathsfbi{P}^{(0) Q} (\boldsymbol{r} - \boldsymbol{r}_c) \, dV,
\label{eq:HotPartCrossTerm}
\end{equation} 
with $b \neq c$.  
Firstly, it can be shown that the dipolar viscosity variations do not affect the result for the $\mathsfbi{Q}^{(1)}$ cross terms up to order $\textit{O}(s^{-2})$ using a similar argument as in the last subsection. 
Secondly, for the calculation of the cross terms up to order $\textit{O}(s^{-2})$, the volumes of both particles can be included in the integration domain without affecting the result. To show this, we consider the Taylor expansion of the integrand around the position of one of the two particles, say particle $b$. In the case $Q = r$, i.e. particle $c \neq b$ is associated to rotation, the normalised strain rate decays with the third power in the inverse distance. With the viscosity variation scaling at most constant in $s$, the resulting error made by including particle $b$ can thus neglected. 
In the case $Q = t$, we have shown in Section \ref{sec:app_4} that the leading order $s^{-2}$ term in the expansion of the contracted normalised strain rates is anti-symmetric under either inversion about the position of particle $b$ for $P = t$ or under $z \to -z$ when $\boldsymbol{s} = (0, 0, s)$ for $P = r$. 
Since both the monopolar viscosity variation around particle $b$ as well as leading order term in the expansion of the monopolar viscosity perturbation around particle $c$ expanded around particle $b$ are symmetric under both of these coordinate transformations, the integral over the volume of particle $b$ vanishes at order $\textit{O}(s^{-2})$. Thus, we can include the volumes of both spheres without affecting the result of the cross terms at our level of approximation. Including the volumes of both particles technically brings into play singularities of the integrand of Eq. \eqref{eq:HotPartCrossTerm} at the particle centres. However, those singularities come with a prefactor which decays quicker than $s^{-2}$, such that we avoid those singularities by Taylor expanding the integrand, expressed in bispherical coordinates, in $s$ and keeping only terms up to order $\textit{O}(s^{-2})$ prior to performing the integration (as also done in Section \ref{sec:app_4}). This integration is performed using Mathematica \citep{Mathematica2020} and yields 
\begin{eqnarray}
&&\mathsfbi{Q}^{(1)tt}_{bc} = \frac{1}{6 \pi a (\eta^{(0)})^2} \left[ -\frac{9 a^2}{8} \left( \eta^{(1) m}_1 + \eta^{(1) m}_2 \right) \frac{\boldsymbol{s} \otimes \boldsymbol{s}}{s^4} \right]  + \textit{O}(s^{-3}), b \neq c, \nonumber \\
&&\mathsfbi{Q}^{(1)tr}_{bc} = \mathsfbi{Q}^{(1) rt}_{bc} = \mathsfbi{Q}^{(1) rr}_{bc} = \textit{O}(s^{-3}) \ \ \forall b \neq c.
\end{eqnarray}

\subsection{Proof of Eq. \eqref{eq:muEqualsQHot}}\label{sec:app_5.3}
For viscosity variations of the form of Eq. \eqref{eq:viscosityPerturbationHotPart}, cross terms as well as terms of the form $\mathsfbi{Q}^{(1) st}_{bb}$ or $\mathsfbi{Q}^{(1) ts}_{bb}$ generally scale at most as $s^{-1}$ or lower, and thus all corrections to $\mathsfbilow{\mu}^{(1) tt}_{bc}$ in Eq. \eqref{eq:muInTermsOfQ} are of the order $\textit{O}(s^{-3})$ and can be neglected. Moreover, terms of the form $\mathsfbi{Q}^{(1) rs}_{bb}$ or $\mathsfbi{Q}^{(1) sr}_{bb}$ are found to be anti-symmetric at leading order $s^0$ in all three indices, similarly to the case of a linear viscosity gradient, and thus their contraction to the symmetric indices of the normalised strain rate vanishes at this order. 
This shows Eq. \eqref{eq:muEqualsQHot}. 

\section{Three-body and higher order effects} \label{sec:app_6}
We prove that for the linear gradient and at the level of approximation chosen, i.e. up to $\textit{O}(s^{-1})$, three-body and higher order effects can be neglected and hydrodynamic interaction thus can be assumed to be pairwise. 
The first-order correction to the mobility matrix is given by \eqref{eq:correctionMobilityGeneral}, where the normalised strain rates $\mathsfbi{B}^{(0) t}_b$ and $\mathsfbi{B}^{(0) t}_c$ need to be expanded as
\begin{equation}
\mathsfbi{B}^{(0) t}_b \equiv \mathsfbi{B}^{(0) t}_b (\boldsymbol{r}) = \mathsfbi{P}^{(0) t} (\boldsymbol{r} - \boldsymbol{r}_b) + \sum_{d \neq b} \mathsfbi{P}^{(0) s} (\boldsymbol{r} - \boldsymbol{r}_d) : \mathsfbi{P}^{(0) t} (\boldsymbol{r}_d - \boldsymbol{r}_b) + \textit{O}(s^{-3}),
\label{eq:expansion1}
\end{equation}
and 
\begin{equation}
\mathsfbi{B}^{(0) t}_c \equiv \mathsfbi{B}^{(0) t}_c (\boldsymbol{r}) = \mathsfbi{P}^{(0) t} (\boldsymbol{r} - \boldsymbol{r}_c) + \sum_{e \neq c} \mathsfbi{P}^{(0) s} (\boldsymbol{r} - \boldsymbol{r}_e) : \mathsfbi{P}^{(0) t} (\boldsymbol{r}_e - \boldsymbol{r}_c) + \textit{O}(s^{-3}).
\label{eq:expansion2}
\end{equation}
The second reflection of a monopolar flow field initially scaling as $s^0$ is a disturbance flow which scales to leading order as $\textit{O}(s^{-5})$ \citep[p. 196]{KimKarilla1991}, and thus such terms can be neglected at our level of approximation. 

Inserting Eqs. \eqref{eq:expansion1} and \eqref{eq:expansion2} into \eqref{eq:correctionMobilityGeneral}, the lowest-order term in $s$ for one particle's self-mobility ($c = b$), which is not captured by pairwise effects, is the product 
\begin{equation}
    \left( \mathsfbi{P}^{(0) s} (\boldsymbol{r} - \boldsymbol{r}_d) : \mathsfbi{P}^{(0) t} (\boldsymbol{r}_d - \boldsymbol{r}_b)\right) \doublecontract \left(\mathsfbi{P}^{(0) s} (\boldsymbol{r} - \boldsymbol{r}_e) : \mathsfbi{P}^{(0) t} (\boldsymbol{r}_e - \boldsymbol{r}_b) \right),
\end{equation}
with $d \neq e$. This term however is a cross term with a prefactor proportional to $s^{-4}$ and can be neglected, as well as all higher-order terms. 

For the interaction between two particles ($b \neq c$), the lowest-order term exceeding pairwise effects is the product
\begin{equation}
    \mathsfbi{P}^{(0) t} (\boldsymbol{r} - \boldsymbol{r}_b) \doublecontract \left( \mathsfbi{P}^{(0) s} (\boldsymbol{r} - \boldsymbol{r}_e) : \mathsfbi{P}^{(0) t} (\boldsymbol{r}_e - \boldsymbol{r}_c) \right),
\end{equation}
with $b \neq e \neq c$, which is a cross term with a prefactor of $\textit{O}(s^{-2})$ and thus can also be neglected. The next higher term is given by products
\begin{equation}
    \left( \mathsfbi{P}^{(0) s} (\boldsymbol{r} - \boldsymbol{r}_d) : \mathsfbi{P}^{(0) t} (\boldsymbol{r}_d - \boldsymbol{r}_b)\right) \doublecontract \left(\mathsfbi{P}^{(0) s} (\boldsymbol{r} - \boldsymbol{r}_e) : \mathsfbi{P}^{(0) t} (\boldsymbol{r}_e - \boldsymbol{r}_c) \right),
\end{equation}
which in the case of $b \neq d = e \neq c$ is a self term, with a prefactor of $\textit{O}(s^{-4})$ however, and thus also can be neglected. 

For hot particles, three-body effects indeed affect the self-mobility of say particles $b$ at order $\textit{O}(s^{-2})$, as the reflection of the temperature of some particle $c$ at some other particle $d$ will alter the viscosity near particle $b$ proportional to $s^{-2}$, and thus cannot be neglected at this level of approximation.

\bibliography{Ziegler202005.bib}

\begin{thebibliography}{29}
\expandafter\ifx\csname natexlab\endcsname\relax\def\natexlab#1{#1}\fi
\def\au#1{#1} \def\ed#1{#1} \def\yr#1{#1}\def\at#1{#1}\def\jt#1{\textit{#1}}
  \def\bt#1{#1}\def\bvol#1{\textbf{#1}} \def\vol#1{#1} \def\pg#1{#1}
  \def\publ#1{#1}\def\arxiv#1{#1}\def\org#1{#1}\def\st#1{\textit{#1}}

\bibitem[Andrade(1930)]{Andrade1930}
{\sc \au{Andrade, E.N. Da~C.}} \yr{1930}  \at{{The Viscosity of Liquids}}.
  \jt{Nature}  \bvol{125},  \pg{309--310}.

\bibitem[B{\"{a}}uerle {\em et~al.\/}(2018)B{\"{a}}uerle, Fischer, Speck \&
  Bechinger]{Bauerle2018}
{\sc \au{B{\"{a}}uerle, Tobias}, \au{Fischer, Andreas}, \au{Speck, Thomas} \&
  \au{Bechinger, Clemens}} \yr{2018}  \at{{Self-organization of active
  particles by quorum sensing rules}}.  \jt{Nature Communications}
  \bvol{9}~(1),  \pg{1--8}.

\bibitem[Buttinoni {\em et~al.\/}(2012)Buttinoni, Volpe, K{\"{u}}mmel, Volpe \&
  Bechinger]{Buttinoni2012}
{\sc \au{Buttinoni, Ivo}, \au{Volpe, Giovanni}, \au{K{\"{u}}mmel, Felix},
  \au{Volpe, Giorgio} \& \au{Bechinger, Clemens}} \yr{2012}  \at{{Active
  Brownian motion tunable by light}}.  \jt{Journal of Physics Condensed Matter}
   \bvol{24}~(28),  \arxiv{arXiv: 1110.2202}.

\bibitem[Dabros(1989)]{Dabros1989}
{\sc \au{Dabros, T.}} \yr{1989}  \at{{Interparticle hydrodynamic interactions
  in deposition processes}}.  \jt{Colloids and Surfaces}  \bvol{39}~(1),
  \pg{127}.

\bibitem[Datt \& Elfring(2019)]{Datt2019a}
{\sc \au{Datt, Charu} \& \au{Elfring, Gwynn~J.}} \yr{2019}  \at{{Active
  Particles in Viscosity Gradients}}.  \jt{Physical Review Letters}
  \bvol{123}~(15),  \pg{158006}.

\bibitem[{De Haan} \& Slater(2013)]{DeHaan2013}
{\sc \au{{De Haan}, Hendrick~W.} \& \au{Slater, Gary~W.}} \yr{2013}
  \at{{Translocation of a polymer through a nanopore across a viscosity
  gradient}}.  \jt{Physical Review E - Statistical, Nonlinear, and Soft Matter
  Physics}  \bvol{87}~(4),  \pg{1--13}.

\bibitem[Dhont(1996)]{Dhont1996}
{\sc \au{Dhont, Jan K~G}} \yr{1996} {\em {An Introduction to Dynamics of
  Colloids}\/}.  \publ{Elsevier}.

\bibitem[Dincer \& Zamfirescu(2015)]{Dincer2015}
{\sc \au{Dincer, Ibrahim} \& \au{Zamfirescu, Calin}} \yr{2015} {\em {Drying
  Phenomena: Theory and Applications}\/}.  \publ{John Wiley \& Sons, Ltd.}

\bibitem[Golestanian \& Ajdari(2008)]{GolestanianAjdari2008}
{\sc \au{Golestanian, Ramin} \& \au{Ajdari, Armand}} \yr{2008}  \at{{Analytic
  results for the three-sphere swimmer at low Reynolds number}}.  \jt{Physical
  Review E - Statistical, Nonlinear, and Soft Matter Physics}  \bvol{77}~(3),
  \pg{036308}.

\bibitem[Gutmann \& Simmons(1952)]{Gutmann1952}
{\sc \au{Gutmann, F.} \& \au{Simmons, L.~M.}} \yr{1952}  \at{{Temperature
  dependence of the viscosity of liquids}}.  \jt{Journal of Applied Physics}
  \bvol{23}~(9),  \pg{977--978}.

\bibitem[Jeffrey \& Onishi(1984)]{Jeffrey1984}
{\sc \au{Jeffrey, D.~J.} \& \au{Onishi, Y.}} \yr{1984}  \at{{Calculation of the
  resistance and mobility functions for two unequal rigid spheres in
  low-Reynolds-number flow}}.  \jt{Journal of Fluid Mechanics}  \bvol{139},
  \pg{261--290}.

\bibitem[Kim \& Karilla(1991)]{KimKarilla1991}
{\sc \au{Kim, Santage} \& \au{Karilla, Seppo~J.}} \yr{1991} {\em
  {Microhydrodynamics : Principles and Selected Applications}\/}.
  \publ{Courier Corporation}.

\bibitem[Kirkwood \& Riseman(1948)]{Kirkwood1948}
{\sc \au{Kirkwood, John~G.} \& \au{Riseman, Jacob}} \yr{1948}  \at{{The
  intrinsic viscosities and diffusion constants of flexible macromolecules in
  solution}}.  \jt{The Journal of Chemical Physics}  \bvol{16}~(6),
  \pg{565--573}.

\bibitem[Kroy {\em et~al.\/}(2016)Kroy, Chakraborty \& Cichos]{Kroy2016}
{\sc \au{Kroy, Klaus}, \au{Chakraborty, Dipanjan} \& \au{Cichos, Frank}}
  \yr{2016}  \at{{Hot microswimmers}}.  \jt{European Physical Journal: Special
  Topics}  \bvol{225}~(11-12),  \pg{2207--2225}.

\bibitem[Laumann \& Zimmermann(2019)]{Laumann2019}
{\sc \au{Laumann, Matthias} \& \au{Zimmermann, Walter}} \yr{2019}
  \at{{Focusing and splitting streams of soft particles in microflows via
  viscosity gradients}}.  \jt{European Physical Journal E}  \bvol{42}~(8),
  \pg{108}.

\bibitem[Leal(2007)]{Leal2007}
{\sc \au{Leal, L.~Gary}} \yr{2007} {\em {Advanced transport phenomena: Fluid
  mechanics and convective transport processes}\/}.  \publ{Cambridge: Cambridge
  University Press}.

\bibitem[Liebchen {\em et~al.\/}(2018)Liebchen, Monderkamp, {Ten Hagen} \&
  L{\"{o}}wen]{Liebchen2018}
{\sc \au{Liebchen, Benno}, \au{Monderkamp, Paul}, \au{{Ten Hagen}, Borge} \&
  \au{L{\"{o}}wen, Hartmut}} \yr{2018}  \at{{Viscotaxis: Microswimmer
  Navigation in Viscosity Gradients}}.  \jt{Physical Review Letters}
  \bvol{120}~(20),  \pg{208002}.

\bibitem[Lu \& Weitz(2013)]{Lu2013}
{\sc \au{Lu, Peter~J.} \& \au{Weitz, David~A.}} \yr{2013}  \at{{Colloidal
  particles: Crystals, glasses, and gels}}.  \jt{Annual Review of Condensed
  Matter Physics}  \bvol{4}~(1),  \pg{217--233}.

\bibitem[Moon \& Spencer(1988)]{Moon1988}
{\sc \au{Moon, P.} \& \au{Spencer, D.E.}} \yr{1988} {\em {Field Theory
  Handbook}\/}.  \publ{Springer Berlin Heidelberg}.

\bibitem[Oppenheimer {\em et~al.\/}(2016)Oppenheimer, Navardi \&
  Stone]{Oppenheimer2016}
{\sc \au{Oppenheimer, Naomi}, \au{Navardi, Shahin} \& \au{Stone, Howard~A.}}
  \yr{2016}  \at{{Motion of a hot particle in viscous fluids}}.  \jt{Physical
  Review Fluids}  \bvol{1}~(1),  \pg{014001}.

\bibitem[Oyama {\em et~al.\/}(2016)Oyama, Molina \& Yamamoto]{Oyama2016}
{\sc \au{Oyama, Norihiro}, \au{Molina, John~Jairo} \& \au{Yamamoto, Ryoichi}}
  \yr{2016}  \at{{Purely hydrodynamic origin for swarming of swimming
  particles}}.  \jt{Physical Review E}  \bvol{93}~(4),  \pg{1--6}.

\bibitem[Pl{\"{u}}isch {\em et~al.\/}(2019)Pl{\"{u}}isch, B{\"{o}}ssenecker,
  Dobler \& Wittemann]{Pluisch2019}
{\sc \au{Pl{\"{u}}isch, Claudia~Simone}, \au{B{\"{o}}ssenecker, Brigitte},
  \au{Dobler, Lukas} \& \au{Wittemann, Alexander}} \yr{2019}  \at{{Zonal rotor
  centrifugation revisited: New horizons in sorting nanoparticles}}.  \jt{RSC
  Advances}  \bvol{9}~(47),  \pg{27549--27559}.

\bibitem[Purcell(1977)]{Purcell1977}
{\sc \au{Purcell, Edward~M}} \yr{1977}  \at{{Life at low Reynolds number}}.
  \jt{Am. Jour. of Phys.}  \bvol{45},  \pg{3--11}.

\bibitem[Qiu \& Mao(2011)]{Qiu2011}
{\sc \au{Qiu, Penghe} \& \au{Mao, Chuanbin}} \yr{2011}  \at{{Viscosity gradient
  as a novel mechanism for the centrifugation-based separation of
  nanoparticles}}.  \jt{Advanced Materials}  \bvol{23}~(42),  \pg{4880--4885}.

\bibitem[Rex \& L{\"{o}}wen(2008)]{Rex2008}
{\sc \au{Rex, M.} \& \au{L{\"{o}}wen, H.}} \yr{2008}  \at{{Influence of
  hydrodynamic interactions on lane formation in oppositely charged driven
  colloids}}.  \jt{European Physical Journal E}  \bvol{26}~(1-2),
  \pg{143--150}.

\bibitem[Shaik \& Elfring(2021)]{Shaik2021}
{\sc \au{Shaik, Vaseem~A} \& \au{Elfring, Gwynn~J}} \yr{2021}
  \at{{Hydrodynamics of active particles in viscosity gradients}}.
  \jt{Physical Review Fluids}  \bvol{6},  \pg{103103}.

\bibitem[Swidsinski {\em et~al.\/}(2007)Swidsinski, Sydora, Doerffel,
  Loening-Baucke, Vaneechoutte, Lupicki, Scholze, Lochs \&
  Dieleman]{Swidsinski2007}
{\sc \au{Swidsinski, Alexander}, \au{Sydora, Beate~C.}, \au{Doerffel, Yvonne},
  \au{Loening-Baucke, Vera}, \au{Vaneechoutte, Mario}, \au{Lupicki, Maryla},
  \au{Scholze, Juergen}, \au{Lochs, Herbert} \& \au{Dieleman, Levinus~A.}}
  \yr{2007}  \at{{Viscosity gradient within the mucus layer determines the
  mucosal barrier function and the spatial organization of the intestinal
  microbiota}}.  \jt{Inflammatory Bowel Diseases}  \bvol{13}~(8),
  \pg{963--970}.

\bibitem[Wilking {\em et~al.\/}(2011)Wilking, Angelini, Seminara, Brenner \&
  Weitz]{Wilking2011}
{\sc \au{Wilking, James~N.}, \au{Angelini, Thomas~E.}, \au{Seminara, Agnese},
  \au{Brenner, Michael~P.} \& \au{Weitz, David~A.}} \yr{2011}  \at{{Biofilms as
  complex fluids}}.  \jt{MRS Bulletin}  \bvol{36}~(5),  \pg{385--391}.

\bibitem[{Wolfram Research Inc.}(2020)]{Mathematica2020}
{\sc \au{{Wolfram Research Inc.}}} \yr{2020} {Mathematica, Version 12.2}.

\end{thebibliography}

\end{document}